\documentstyle[12pt]{article}
\newcommand\be{\begin{equation}}
\newcommand\ee{\end{equation}}
\newcommand\ba{\begin{eqnarray}}
\newcommand\ea{\end{eqnarray}}
\newcommand\ve{\varepsilon}
\newcommand\la{\langle}
\newcommand\r{\rangle}
\begin{document}

\title{\vskip -2cm\hfill {\rm \normalsize{HD--TVP--94--16}}
\\
\vskip 2cm
How sharp is the chiral crossover phenomenon for realistic meson masses?}
\author{Hildegard Meyer--Ortmanns\thanks{e-mail address: ort@dhdmpi5.bitnet}\\
and\\
Bernd-Jochen Schaefer\thanks{e-mail address:
schaefer@hybrid.tphys.uni-heidelberg.de}
\\\\
Institut f\"ur Theoret. Physik,\\Universit\"at Heidelberg\\
Philosophenweg 19\\ D-69120 Heidelberg, F.R.G.}

\date{ }

\vspace{1.0cm}

\setlength{\baselineskip}{18pt}

\maketitle

\begin{abstract}
\setlength{\baselineskip}{18pt}
\noindent
The mass dependence of the chiral phase transition is studied in the linear
$SU(3)\times SU(3)$ sigma-model to leading order in a $1/N_f$-expansion, $N_f$
denoting the number of flavours.
For realistic meson masses we find a smooth crossover between $T\sim181.5$ to
192.6~[MeV]. The crossover looks more rapid in the light quark condensate than
in
thermodynamic quantities like the energy and entropy densities.
The change in the light quark condensate in this temperature interval
is $\sim$~50\% of the zero-temperature condensate value,
while the entropy density increases by
($5.5\pm0.8)\cdot10^{-3}$~[GeV$^3$]. Since the numerical error is
particularly large
in this
region, we cannot rule out a finite latent heat smaller than 0.2~[GeV/fm$^3$].
The chiral transition is washed out for an average pseudoscalar meson octet
mass of
203~[MeV]. This gives an upper bound on the first-order transition region in
the meson
mass parameter space. The corresponding ratio of critical to realistic light
current
quark masses $m^{crit}_{u,d}/m_{u,d}$ is estimated as $0.26\pm0.08$.
This result is by an order of magnitude larger than the corresponding
mean-field
value. Therefore the realistic quark/meson masses seem to lie less deeply in
the
crossover region than it is suggested by a mean-field calculation.
\end{abstract} \newpage \parskip12pt

\section{Introduction}
Spontaneous symmetry breaking is frequently used as an ansatz to explain the
driving
force of finite temperature phase transitions in QCD. The symmetries refer to
certain
limiting cases of the QCD-Lagrangian. In the limit of infinite quark masses and
in the
case of three colours finite temperature QCD is invariant under
$Z(3)$-transformations ($Z(3)$ is the center of $SU(3)$). The spontaneous
breaking of
$Z(3)$ is made responsible for the phase transition from the confinement phase
at low
temperatures to the deconfinement phase at high temperatures. In the other
extreme
case of $N_f$ vanishing current quark masses QCD is invariant under
$SU(N_f)\times
SU(N_f)$ chiral transformations. The restoration of the spontaneously broken
chiral
symmetry at high temperatures is said to drive the chiral phase transition of
QCD. For
the chiral limit the renormalization group analysis of Pisarski and Wilczek
\cite{pis}
serves as a guideline for conjectures about the order of the chiral phase
transition.
Similar studies have been performed by Svetitsky and Yaffe \cite{sve} in the
other
extreme case of pure gauge theory or infinite quark masses.

In reality the current quark masses are neither infinite nor zero. The masses
of the
charm, bottom and top quarks are large compared to the scale of the critical
temperature $T_c$ of QCD, which lies between 150 and 250~[MeV]. Thus it seems
to be
justified to treat them as infinite in thermodynamic investigations. The
(renormalization group invariant current) masses of the up and down quarks are
small
compared to $T_c,\ m_u=7.6\pm2.2$~[MeV], $m_d=13.3\pm3.9$~[MeV] for
$\Lambda=100$~[MeV],
where $\Lambda$ refers to the $\overline{MS}$ scheme with 3
flavours~\cite{gas}.
 At a first glance it seems to be well justified to set them simply to
zero, although more careful investigations in the framework of chiral
perturbation
theory give us a warning to neglect their influence on the chiral transition
\cite{ger}. A particular role is played by the strange quark. Its mass
($m_s\sim205\pm50$~[MeV]) is neither small nor large, but just of the order of
the scale
which is set by $T_c$. To get an idea about the influence of finite quark
masses on the
phase structure of QCD it is instructive to consult statistical physics. The
phase
transition from a liquid to a gas is of first order below some critical value
of the
pressure. For a critical value of the pressure it becomes of second order.
Above the
critical strength it turns into a smooth crossover between the liquid and the
gas
phase. Similar effects are known from ferromagnets under the influence of an
external
magnetic field. In an $O(N)$-ferromagnet an arbitrarily small magnetic field is
sufficient to turn the second order transition with an infinite correlation
length
into a crossover without a diverging correlation length. In close analogy to
our
subsequent considerations we should also mention the three-dimensional
$Z(3)$-Potts
model. The spin variables of this model can take three values at each lattice
site. For
a vanishing external magnetic field the Potts model is known to have a first
order
phase transition at finite temperature. For a critical field strength it
becomes of
second order and disappears for even stronger external fields.

The conjugate variables are the pressure and the specific volume in the
liquid/gas
system, the external magnetic field and the magnetization for the ferromagnet.
The
analogous pair in QCD are the current quark masses and the quark
condensates as the associated order parameters for the chiral transition (on
the
quark level). On the mesonic level we have external fields, which can be
related to the
quark masses, and mesonic condensates. The comparison between the statistical
systems
and QCD goes beyond a formal analogy. In an $SU(3)$-lattice gauge theory it has
been
shown by DeGrand and DeTar \cite{deg} and Banks and Ukawa \cite{ban} that
dynamical
fermions on the QCD level induce an external field coupled linearly to a spin
field in
an effective $Z(3)$-spin model. The spin model has been {\it derived} from QCD
in a
strong coupling expansion at high temperatures. The external field strength was
expressed in terms of the hopping parameter of the lattice-QCD Lagrangian.

\vspace{24pt}

In this paper, we consider the $SU(3)\times SU(3)$ linear sigma-model as an
effective
model for chiral symmetry restoration. We include two external fields to
account for
the finite quark masses on the mesonic level. The action is constructed in
terms of
the chiral order parameter field, which is a matrix of mesonic condensates. In
the
special case of one external field and one mesonic condensate, the action of
the
sigma-model takes the form of Landau's free energy functional for a liquid/gas
system
(cf. section~5). Thus we expect the same qualitative features as we know from
the
liquid/gas system. Beyond certain critical values of the meson masses the
chiral
transition should be washed out and turn into a crossover phenomenon with
smooth
changes in the condensates, energy density and entropy density. Therefore the
main
question is, as to whether the realistic quark masses are too large for the
chiral
transition and too small for the deconfinement transition to persist.

Numerous numerical simulations in lattice QCD have investigated the phase
structure
of QCD. Results on the mass dependence of the finite temperature transitions
were
summarized in the famous Columbia-plot \cite{col}, cf. Fig.~1.
\begin{figure}[thb]
\vspace{7.0cm}
\caption{Columbia plot [7]. Generic phase diagram, partly conjectural, for 2
and
3 flavours. Solid lines indicate supposed 2nd order transitions, shaded areas
1st order
transitions. Solid circles correspond to mass parameters, which lead to 1st
order
transitions, triangles to mass parameters, which lead to crossover phenomena.
The
open circle locates the suggested physical mass point.}
\end{figure}
Indicated are the presence or absence of QCD transitions as a function of the
bare
quark masses $m_s\cdot a$ and $m_{u,d}\cdot a$ in lattice units, $a$ denotes
the
lattice constant. The solid line stands for second order transitions, the
shaded
areas enclose mass values leading to first order transitions. Solid circles
correspond to mass values leading to 1st order transitions, e.g. the point with
3
degenerate bare quark masses $m_ua=m_da=m_sa=0.025$. Triangles indicate mass
combinations which lead to a crossover phenomenon. In particular, the mass
point is
included with two light ($m_ua=m_da=0.025$) and one heavier flavour
($m_sa=0.1$).
This mass point comes closest to the realistic quark masses with a ratio of
$m_s/m_{u,d}\sim20$. The ratio between the critical quark mass $m_u^{crit}$
which
lies on the first order transition boundary to the realistic quark mass $m_u$
is
about 0.5 if the ratio $m_u^{crit}a/m_s^{crit}a=m_ua/m_sa$ is kept fixed.
Fig.~1
suggests that the masspoint with realistic quark masses is not too far from the
second order transition line, although the precise values of the critical
strange
quark masses, where the first order transitions turn into second order
transitions,
are still an open question.

Thus the conclusion from the lattice is that there is no true chiral phase
transition for experimental quark masses. Depending on the degree of smoothness
of
the chiral transition this conclusion  can have far reaching consequences in
view of
measurable effects in heavy-ion collisions. Therefore we spend some comments on
the
reliability of the lattice result.

Monte Carlo results from lattice calculations are in general affected by
artifacts
due to the finite IR-cutoff (i.e. the finite volume), the UV-cutoff (i.e. the
finite
lattice constant) and finite bare quark masses (it is unavoidable to {\it
extrapolate} to the chiral limit). Applied to the masspoint with two light and
one
heavier flavour ($m_{u,d}a=0.025,m_sa=0.1$) of the Columbia plot \cite{col}
there
are some warnings which should be kept in mind. The lattice extension in
imaginary
time direction was chosen as $N_\tau=4$. Even without dynamical fermions
$N_\tau$
should be larger than 10 to find a critical bare coupling close to the
continuum
limit. One manifestation of the vicinity to the strong coupling regime are the
masses of flavour partners which should be degenerate in the continuum limit.
The
ratio of two $K$-meson masses is still about 2 \cite{col}. The effect of
heavier
flavour partners on the order of the chiral transition is difficult to control.

Another subtle effect of the finite UV-cutoff is a change in the effective
symmetry
as a function of the bare coupling if staggered fermions are used \cite{boy}.
For
four continuum flavours the chiral transition is of 2nd order in the strong
coupling
limit, but of 1st order close to the continuum region. For two flavours the
critical
indices of the second order transition change between the strong and weak
coupling
region \cite{kar}. For three flavours one can speculate, whether the crossover
itself is a strong coupling feature rather than continuum physics.

A further obstacle in the staggered fermion formulation applies in particular
to the
case of three flavours. The effective fermionic action which projects on three
flavours must be regarded as a prescription rather than being derived from the
staggered fermion action \cite{col}. A derivation could only lead to integer
multiples of four continuum flavours in the effective action. The error due to
this
prescription seems to be difficult to control.

A bulk ``chiral'' transition has been identified for 8 flavours \cite{chr} (for
bulk transitions the transition itself is a mere lattice artifact). Precursors
of
this bulk transition for a smaller number of flavours can be superimposed on
the
finite $T$-chiral transition and influence its strength.

If the bare coupling lies in the strong coupling regime, the conversion of bare
masses in lattice units into masses in [MeV] is not unique. Last but not least
it is
not clear whether the ratio of $m_{u,d}a/m_sa$ is unaffected by multiplicative
renormalizations (it may be that $m^{ren}_{u,d}/m_s^{ren}\not=m_{u,d}a/m_sa$).
Hence
the very location of the physical mass point in the $(m_{u,d}a,m_sa)$ diagram
is
questionable.\\
These warnings should suffice as arguments in favour of alternative
approaches which include different approximations to study the phase structure
of QCD.

In \cite{mey} a systematic study of the mass dependence of the chiral
transition
has been proposed in the framework of {\it effective models} for QCD. In a
first
paper \cite{mey} a bound on the first order transition region was given for
the $SU(3)$-symmetric case. The loose bound on the (average) pseudoscalar octet
mass
was 100~[MeV] for a sigma meson mass between 600 and 950~[MeV]. The qualitative
conclusion
was that realistic meson masses lie deeply in the crossover region. The results
were
obtained in a large $N_f$-expansion under the omission of $n\not=0$-Matsubara
frequencies.

In this paper we will give a more stringent bound on the maximal pseudoscalar
octet
mass, for which the transition is still of first order.
The large $N_f$-expansion seems to be a good starting point for the
$SU(3)\times SU(3)$ linear sigma-model, which reduces to an $O(18)$-model in a
special
case of two vanishing couplings, see section 3 below.
Therefore we still use the large $N_f$-expansion, but include also the
$n\not=0$-Matsubara frequencies. The omission of the $n\not=0$-modes seems to
be
justified only for high temperatures or in case of a second order phase
transition.
The
distance (in mass parameter space) between experimental meson masses and
critical
meson masses, for which the first order chiral transition turns into a
crossover
phenomenon, is more than a quantity of academic interest. Its physical
relevance can be seen as follows. Assume that the physical masses are not
identical but very close to the critical mass values where the chiral
transition is
of second order. In such a case the chances are good to see some remnant in
heavy-ion
collisions of a
diverging correlation length for critical mass values, i.e. a rather large
correlation length for realistic masses.\\
One manifestation of a large correlation length has been supposed to be large
clusters of charged or neutral pions which are aligned in isospin space
\cite{wil}. The correlation volume of a cluster should be large enough that the
number of emitted pions is sufficient for the detector to resolve the cluster
structure. Intermittent behaviour has been proposed as another manifestation of
a
second order transition \cite{bia}. In this case one would observe some kind of
self-similarity phenomenon if the bin size of rapidity intervals is made
finer and finer.\\
Also the vicinity to the first order transition region could leave observable
effects if the crossover for realistic masses is still sharp enough. A sharp
crossover is associated with a rapid change in the condensates and their
induced
masses, and/or a rapid change of entropy in a small temperature interval.
Average
transverse momentum distributions of charged particles could be flattened as a
function of the multiplicity $dN/dy$ of final state particles in a given
rapidity interval \cite{bla}. Pronounced fluctuations in the particle
multiplicities would show up if the crossover is strong enough to induce
deflagration
processes during the phase conversion \cite{hov}. True singularities of a first
or
second order transition will be anyway rounded in real experiments due to the
finite
volume.

It may be academic to ask as to whether chiral symmetry restoration in the
infinite
volume limit proceeds via a weakly first order transition or a smooth crossover
phenomenon. It is certainly more sensible to pose the question in the following
way: Is the gap in entropy densities in the transition/crossover region
sufficient
to induce multiplicity fluctuations in the observed pion yield, lying clearly
above
the statistical noise? In this paper, we therefore try to find a quantitative
measure for how sharp the chiral crossover phenomenon is for realistic meson
masses.

The paper is organized as follows. In section 2 we present the tree level
parametrization of the $SU(3)\times SU(3)$-linear sigma-model at zero
temperature.
We give a prescription to translate the meson condensates and meson masses to
quark
condensates and current quark masses. In section 3 we summarize the essence of
our approach. The mesonic self-interaction is treated to leading order in an
expansion in the number of quark flavours $N_f$. On the mesonic level we have 9
scalar plus 9 pseudoscalar mesons. The $SU(3)\times SU(3)$ sigma-model reduces
to
an $O(18)$-model in a certain limit. Thus the leading order of a
$1/N$-expansion
should be a good starting point. The thermodynamic effective potential is
evaluated
in a high-temperature expansion and in a fully numerical approach. The
numerical
approach is certainly more appropriate to the phase transition region. To
discuss
the mass dependence of the order of the chiral transition we distinguish three
regions in mass parameter space $(m_\pi,m_K,\ldots)$ and $(m_{u,d},m_s)$: the
chiral limit (section 4),
several mass points on the first order transition boundary, so-called critical
mass
points (section 5) and realistic meson masses (section 6), which come close to
the
experimental masses. In the chiral limit we fix the couplings of the tree level
parametrization. The quantitative discrepancy between the high-temperature
expansion and the fully numerical evaluation will be revealed. We
calculate the barrier height between the coexisting minima of the effective
potential as one measure for the strength of the first order transition.\\
Values of critical meson masses are determined in section 5 for three special
cases: the $SU(3)$-symmetric case with finite, but degenerate pseudoscalar
octet
masses, the case with a realistic mass splitting between the masses of the
pseudoscalar octet (the ratios of meson masses with and without strangeness are
kept fixed to their experimental values), and the case of vanishing strange
quark
mass $m_s$. The critical meson masses are calculated both in a mean-field
approximation and in the large $N_f$-expansion.\\
Section 6 deals with meson masses which are rather close to the experimental
values. We describe the crossover phenomenon in the meson and quark condensates
as
a function of temperature. The crossover is also manifest in thermodynamic
quantities like the energy density $\varepsilon$, the entropy density $s$, and
the
pressure $p$. We derive $\varepsilon,s$ and $p$ from the partition function of
the
sigma-model in a saddle-point approximation. Upper bounds on a finite latent
heat
during the chiral transition are predicted. The temperature dependence of
effective
masses which enter the calculation of the effective potential will be
displayed. In
section 7 we summarize our results and draw some conclusions in view of
heavy-ion
experiments.

\section{Tree level parametrization of the $SU(3)\times SU(3)$ linear
sigma-model}

The Euclidean Lagrangian density of the $SU(3)\times SU(3)$-linear sigma-model
is
given as
\ba
{\cal L}_{Eucl}&=&\frac{1}{2}{\rm Tr}(\partial_\mu M\partial_\mu
M^+)-\frac{1}{2}
\mu^2_0{\rm Tr}MM^+ +g(\det M+\det M^+)+ \nonumber\\
&+&f_1({\rm Tr}MM^+)^2+f_2{\rm
Tr}(MM^+)^2-\varepsilon_0\sigma_0-\varepsilon_8\sigma_8, \label{aa}
\ea
where the $(3\times3)$-matrix field $M(x)$ is written as
\be
M=\frac{1}{\sqrt2}\sum^8_{\ell=0}(\sigma_\ell+i\ \pi_\ell)\lambda_\ell.
\label{ab}\ee
Here $\sigma_\ell$ and $\pi_\ell$ denote the nonets of scalar and pseudoscalar
mesons, respectively, $\lambda_\ell\ (\ell=1,...,8)$ are the Gell-Mann
matrices,
$\lambda_0=\sqrt{\frac{2}{3}}\cdot{\rm diag}(1,1,1)$. The chiral $SU(3)\times
SU(3)$-symmetry is explicitly broken by the term
$(-\varepsilon_0\sigma_0-\varepsilon_8\sigma_8)$, which is linear in the
external
fields $\varepsilon_0,\varepsilon_8$. A non-vanishing value of $\varepsilon_0$
gives a common mass value to the octet of pseudoscalar Goldstone bosons
$m_\pi,m_k,m_\eta$. When also $\varepsilon_8\not=0$, it can be adjusted such
that
it leads to a realistic mass splitting inside the (pseudo)scalar meson octets.

The Lagrangian (\ref{aa}) appears as a natural candidate for an effective
model,
which is designed to describe the phenomenon of chiral symmetry restoration.
The
action $S=\int d^3xd\tau{\cal L}(x)$ with ${\cal L}$ of Eq.~(\ref{aa}) may be
regarded as an effective action for QCD, constructed in terms of a chiral order
parameter field $M$ for the chiral transition. As order parameters for the
chiral
transition we choose the meson condensates $\langle\sigma_0\rangle$ and
$\langle\sigma_8\rangle$. The expectation value of $M$ is then parametrized in
terms of $\langle\sigma_0\rangle,\langle\sigma_8\rangle$ according to
\be
\label{ac}
\langle M\rangle={\rm
diag}\frac{1}{\sqrt3}\left[\langle\sigma_0\rangle+\frac{1}{\sqrt2}
\langle\sigma_8\rangle,\langle\sigma_0\rangle+\frac{1}{\sqrt2}
\langle\sigma_8\rangle,\langle\sigma_0\rangle-\sqrt2\langle\sigma_8\rangle\right].\ee
The construction of an action $S$ in terms of an order parameter field is a
concept
in close analogy to Landau's free energy functional ${\cal F}$ in terms of an
order
parameter field, ${\cal F}$ coincides with $S$ in the mean-field approximation.
Quartic terms in $M$ have to be introduced in Eq.~(\ref{aa}) to allow for the
possibility of spontaneous symmetry breaking. For
$g=0=\varepsilon_0=\varepsilon_8$
the Lagrangian is still invariant under $U(3)\times U(3)$ transformations. For
$N_f\geq3$ there are two independent quartic terms, parametrized with
coefficients
$f_1$ and $f_2$. To account for a realistic $(\eta,\eta')$-mass splitting, a
$\det$-term with an ``instanton''-coupling $g$ has to be included as well. It
reduces
the symmetry of ${\cal L}$ to $SU(3)\times SU(3)$ if $\ve_0=0=\ve_8$. Finally,
the external field terms which are linear in $M$ are the most simple choice for
an
explicit symmetry breaking accounting for the small, but finite masses of the
Goldstone octet $(m_\pi,m_K,m_\eta)$. Thus one arrives at the $SU(3)\times
SU(3)$
linear sigma-model in a natural way if one is interested in the limited aspect
of
chiral symmetry restoration.

It remains to fix the Lagrangian parameters $\mu_0^2,f_1,f_2,g,\ve_0,\ve_8$
from an
experimental input. The choice of the input and the way of parametrization are
in
no way unique, cf. \cite{cha,gav,mey}. Since the pseudoscalar meson
masses  are experimentally well known, $m_\pi,m_k,m_\eta,m_{\eta'}$ and $f_\pi$
have been used in \cite{mey} to fix $\mu_0^2,f_1,f_2,g,\ve_0$ and $\ve_8$. In
addition, the mass of the $\sigma'$-meson has been treated as input parameter
and
varied between 600 and 950~[MeV] to solve for $\mu_0^2$ and $f_1$, which occur
in the
(pseudo)scalar meson masses in the combination
$(-\mu^2_0+4f_1(\sigma^2_0+\sigma^2_8))$. It is worth noting that the order
parameters $\la\sigma_0\r,\la\sigma_8\r$ can be determined without knowing
$\mu^2_0$
and $f_1$ separately. The equations for  $\mu_0^2,f_1,f_2,g,\ve_0,\ve_8$ do not
admit solutions for an arbitrary choice of  $m_\pi,m_k,m_\eta,m_{\eta'},f_\pi$
and
$m_\sigma$. This was the reason why the input masses which have been actually
used as
input in \cite{mey} were slightly deviating from the experimental values if
$m_{\sigma_{\eta'}}$ was chosen as 600 or 950~[MeV]. In \cite{met} the
experimental
values could be used for $m_\pi,m_k,m_\eta,m_{\eta'}, f_\pi$ on the price that
$m_{\sigma_{\eta'}}$ was used as input with 1400~[MeV].

In this paper, our interest goes beyond the point with (almost) experimental
pseudoscalar meson masses.
Naturally, the question arises why we are not satisfied with a parametrization
that
reproduces the experimental values for the (pseudo)scalar meson masses, but
want to
tune some parameters to unphysical, unrealistic mass values as well. There are
two reasons for that. The first one is to check how stable our results about
the
phase structure are under a slight change of the mass input. The second one is
to get
a quantitative measure for the distance (in mass parameter space) between the
realistic and the critical meson masses, cf. section 5 below.
As we focus on the aspect of the mass sensitivity, we have to find a
prescription how to tune the masses in the
high-dimensional meson mass parameter space. On the quark level the mass
parameter
space is only two-dimensional, the two parameters are $m_{u,d}$ and $m_s$.
Not only the parametrization, but also
the tuning in the space of meson masses is by far not unique. The idea now is
to
parametrize the (pseudo)scalar meson masses by two parameters like the quark
masses, the external fields $\ve_0$ and $\ve_8$.

A relation between $m_{u,d},m_s$ and $\ve_0,\ve_8$ is obtained by identifying
terms
of the Lagrangian on the mesonic and on the quark level which transform
identically
under $SU(3)\times SU(3)$. We have
\ba
\label{ad}(-\ve_0\sigma_0-\ve_8\sigma_8)& & \mbox{on the mesonic level},\\
\label{ae}+(m_u\bar uu+m_d\bar dd+m_s\bar ss)& & \mbox{on the quark level}.
\ea
Thus we find
\ba \label{af}
-\ve_0&=&\alpha(2\hat m+m_s)\nonumber\\
-\ve_8&=&\beta(\hat m-m_s). \ea
Here $\alpha$ and $\beta$ are constants. They can be fixed from the known
values of
$\ve_0,\ve_8,m_{u,d}$ and $m_s$ under realistic conditions. Realistic meson
masses
are obtained for $\ve_0=0.0265~{\rm [GeV]}^3,\ve_8=-0.0345~{\rm [GeV]}^3$, see
below.
The values for ``realistic'' current quark masses are taken from \cite{nar},
$\hat
m\equiv(m_u+m_d)/2=11.25\pm1.45$~[MeV], $m_s=205\pm50$~[MeV]. For $\alpha$ and
$\beta$ we then
obtain \be \label{ag}
\alpha=-0.1164\ {\rm [GeV]}^2, \qquad \beta=-0.1780 {\rm [GeV]}^2 . \ee
Thus a variation in $(\ve_0,\ve_8)$ can be mapped onto a variation of $(\hat
m,m_s)$.

Next we have to find a mapping between $(\ve_0,\ve_8)$ and the (pseudo)scalar
meson
masses. As it is possible to explain the variety of (pseudo)scalar meson masses
on
the basis of two quark masses $m_{u,d}$ and $m_s$, it should be similarly
possible
to reach any point in an $(m_\pi,m_{K,\ldots})$-diagram by a variation of
$\ve_0$ and
$\ve_8$. Thus we start with the parametrization of the chiral limit
$\mu^2_0,f_1,f_2,g,\ve_0=0,\ve_8=0$, keep the couplings $\mu_0^2,f_1,f_2,g$
fixed
and switch on $\ve_0\not=0,\ve_8\not=0$. The $SU(3)$-symmetric case with
finite, but
degenerate pseudoscalar meson masses is obtained for $\ve_0\not=0,\ve_8=0$. The
mass point with realistic meson masses which come close to their experimental
values, is obtained for $\epsilon_0=0.0265$~[GeV]$^3,\ve_8=-0.0345$~[GeV]$^3$.
(In
principle, there is no obstacle to further optimize the values of $\ve_0,\ve_8$
such that they do reproduce the experimental mass values). For the chiral
values
$\mu^2_0,f_1,f_2,g$ and a certain choice for $\ve_0$ and $\ve_8$ we first
determine
$\langle\sigma_0\rangle$ and $\langle\sigma_8\rangle$, the condensates at zero
temperature, as zeroes of (\ref{ah}),(\ref{ai}) in $\sigma_0$ and $\sigma_8$
\ba
\ve_0+\mu^2_0\sigma_0-\frac{g}{\sqrt3}(2\sigma^2_0-\sigma^2_8)+\frac{2\sqrt2}{3}
f_2\sigma^3_8-4(f_1+\frac{f_2}{3})\sigma^3_0
-4(f_1+f_2)\sigma_0\sigma_8^2=0&&\nonumber \\ \label{ah}\\
\ve_8+\mu^2_0\sigma_8+\sqrt{\frac{2}{3}}g(\sigma^2_8+\sqrt2\sigma_0\sigma_8)+
2\sqrt2f_2\sigma_0\sigma^2_8-4(f_1+\frac{f_2}{2})\sigma^3_8
-4(f_1+f_2)\sigma^2_0\sigma_8 =0.&&\nonumber\\ \nonumber\\
{}
\label{ai}
\ea
Eqs.~(\ref{ah},\ref{ai}) are the equations of motion for constant background
fields
$\sigma_0,\sigma_8$. The pseudoscalar meson masses are then given as
\ba \label{aj}
m^2_\pi&=&-\mu^2_0+(4f_1+\frac{4}{3}f_2)\sigma^2_0+(4f_1+\frac{2}{3}f_2)\sigma^2_8+
\frac{4}{3}\sqrt2f_2\sigma_0\sigma_8+\frac{2g}{\sqrt3}\sigma_0-2\sqrt{\frac{2}{3}}g
\sigma_8\nonumber\\
m^2_k&=&-\mu^2_0+(4f_1+\frac{4}{3}f_2)\sigma^2_0+(4f_1+\frac{14}{3}f_2)\sigma^2_8
-\frac{2}{3}\sqrt2f_2\sigma_0\sigma_8+\frac{2}{\sqrt3}g\sigma_0+\sqrt{\frac{2}{3}}
g\sigma_8\nonumber\\
m^2_{\eta_{00}}&=&-\mu^2_0+(4f_1+\frac{4}{3}f_2)\sigma^2_0+(4f_1+\frac{4}{3}f_2)
\sigma^2_8-\frac{4}{\sqrt3}g\sigma_0\nonumber\\
m^2_{\eta_{88}}&=&-\mu^2_0+(4f_1+\frac{4}{3}f_2)\sigma^2_0+(4f_1+2f_2)\sigma^2_8
-\frac{4}{3}\sqrt2f_2\sigma_0\sigma_8+\frac{2}{\sqrt3}g\sigma_0+2\sqrt{\frac{2}{3}}g
\sigma_8\nonumber\\
m^2_{\eta_{08}}&=&\sigma_8\left[
\frac{8}{3}f_2\sigma_0-\frac{2}{3}\sqrt2f_2\sigma_8+
\frac{2}{\sqrt3}g\right]\nonumber\\
m^2_\eta&=&\frac{1}{2}(m^2_{\eta_{00}}+m^2_{\eta_{88}}-\sqrt{(m^2_{\eta_{00}}-
m^2_{\eta_{88}})^2+4(m^2_{\eta_{08}})^2})\nonumber\\
m^2_{\eta'}&=&\frac{1}{2}(m^2_{\eta_{00}}+m^2_{\eta_{88}}+\sqrt{(m^2_{\eta_{00}}-
m^2_{\eta_{88}})^2+4(m^2_{\eta_{08}})^2}).
\ea
The scalar meson masses follow from
\ba \label{ak}
m^2_{\sigma_{\pi}}&=&-\mu^2_0+(4f_1+4f_2)\sigma^2_0+(4f_1+2f_2)\sigma^2_8+
\frac{8}{\sqrt2}f_2\sigma_0\sigma_8-\frac{2}{\sqrt3}g\sigma_0+2\sqrt{\frac{2}{3}}
g\sigma_8\nonumber\\
m^2_{\sigma_k}&=&-\mu^2_0+(4f_1+4f_2)\sigma^2_0+(4f_1+2f_2)\sigma^2_8-\frac{4}{\sqrt2}
f_2\sigma_0\sigma_8-\frac{2}{\sqrt3}g\sigma_0-\sqrt{\frac{2}{3}}g\sigma_8\nonumber\\
m^2_{\sigma_{00}}&=&-\mu^2_0+(12f_1+4f_2)\sigma^2_0+4(f_1+f_2)\sigma^8+
\frac{4g}{\sqrt3}\sigma_0\nonumber\\
m^2_{\sigma_{88}}&=&-\mu^2_0+4(f_1+f_2)\sigma^2_0+6(2f_1+f_2)\sigma^2_8-4\cdot\sqrt2
f_2\sigma_0\sigma_8-\frac{2g}{\sqrt3}\sigma_0-2\sqrt{\frac{2}{3}}g\sigma_8\nonumber\\
m^2_{\sigma_{08}}&=&\sigma_8[8(f_1+f_2)\sigma_0-\frac{4}{\sqrt2}f_2\sigma_8-\frac{2}
{\sqrt3}g]\nonumber\\
m^2_{\sigma_{\eta}}&=&\frac{1}{2}(m^2_{\sigma_{00}}+m^2_{\sigma_{88}}+
\sqrt{(m^2_{\sigma_{00}}-
m^2_{\sigma_{88}})^2+4(m^2_{\sigma_{08}})^2})\nonumber\\
m^2_{\sigma_{\eta'}}&=&\frac{1}{2}(m^2_{\sigma_{00}}+m^2_{\sigma_{88}}-
\sqrt{(m^2_{\sigma_{00}}-
m^2_{\sigma_{88}})^2+4(m^2_{\sigma_{08}})^2}),
\ea
where $\sigma_0,\sigma_8$ should be replaced by $\langle\sigma_0\rangle,\langle
\sigma_8\rangle$, respectively, everywhere
in Eqs.~($10,11$) for physical masses. The resulting masses together
with the input parameters are listed in Table~1 of section~6.

To facilitate a comparison with other work on the chiral phase transition, it
remains
to translate the results for the meson condensates to the quark level. If we
identify
terms in the Lagrangians on the quark level and on the mesonic level (cf.
Eqs.~(\ref{ad}) and (\ref{ae})), we find
\ba
\frac{1}{3}(2\hat m+m_s)(2\bar qq+\bar ss)&=&-\ve_0\sigma_0 \label{al}\\
\frac{2}{3}(\hat m-m_s)(\bar qq-\bar ss)&=&-\ve_8\sigma_8 \label{am}
\ea
with $\bar qq =\frac{1}{2}( \bar uu + \bar dd )$,
leading to
\ba \label{an}
\langle\bar qq\rangle&=&\frac{-\ve_0}{2\hat m+m_s}\langle\sigma_0\rangle+
\frac{-\ve_8}{2(\hat m-m_s)}\langle\sigma_8\rangle\nonumber\\
\langle\bar ss\rangle&=&\frac{-\ve_0}{2\hat m+m_s}\langle\sigma_0\rangle+
\frac{+\ve_8}{\hat m-m_s}\langle\sigma_8\rangle.
\ea
The coefficients are just
proportional to $\alpha$ and $\pm \beta$, as a comparison with
Eq.~(\ref{af}) shows, $\alpha$ and $\beta$ have been determined in
Eq.~(\ref{ag}).
Later we take the relations (\ref{an}) as temperature independent and
substitute
$\langle\sigma_0\rangle(T), \langle\sigma_8\rangle(T)$ for the corresponding
condensates at zero
temperature. This assumption is consistent with our approach. We also treat the
couplings $\mu_0^2,f_1,f_2,g$ of the Lagrangian as temperature independent.
Therefore
the symmetry of ${\cal L}$ remains unchanged under an increase of $T$. The
symmetry
was on the basis of the identification which has led to Eqs.~(\ref{an}). (The
assumption of temperature independent couplings may not be justified in the
vicinity
of the transition region.) In the next section we will outline how to calculate
$\langle\sigma_0\rangle(T),\langle\sigma_8\rangle(T)$.

\section{Large $N_f$-expansion}

In an earlier calculation the linear $SU(3)\times SU(3)$ sigma-model has been
considered in a mean-field approximation \cite{gol}. Recently, Gavin, Goksch
and
Pisarski \cite{gav} have tried to localize the first order transition boundary
in an
$(m_{u,d},m_s)$-mass diagram in a {\it mean-field} calculation. The famous
renormalization group analysis of Pisarski and Wilczek \cite{pis} applied to
the
linear sigma-model in the {\it chiral limit}. Frei and Patk\'os \cite{fre} were
the
first to apply a saddle-point approximation to the partition function of the
sigma-model. Their investigations were also restricted to the chiral limit.
Meyer-Ortmanns, Patk\'os and Pirner \cite{mey} have extended the approach of
Frei and
Patk\'os to finite meson masses. In \cite{mey} only the zero-Matsubara
frequencies
were kept. When the imaginary time dependence of the fields or, equivalently,
the
$n\not=0$-Matsubara frequencies are dropped, it results in a dimensional
reduction of the
four-dimensional theory to an effective three-dimensional theory. In general,
such a
reduction can be justified in the high-$T$-limit or for an anticipated second
order
phase transition. In both cases the ratio of $\beta/\xi$ is negligible ($\beta$
is the
inverse temperature and $\xi$ denotes the largest correlation length of the
system).
The $n\not=0$-Matsubara modes were included in \cite{met}.
In this paper, we follow the same approach as in \cite{met}, but extend the
work to
study the aspect of the mass sensitivity of the chiral transition. Further
differences in details of \cite{met} and the present paper will be mentioned
below. In
the following, we summarize the essence of the large-$N_f$-approach to make the
paper
self-contained.

The temperature-dependent order parameters are the meson condensates
$\la\sigma_0\r(T)$ and $\la\sigma_8\r(T)$. They are related to the light and
strange
quark condensates according to Eqs.~(\ref{an}). The values of
$\la\sigma_0\r(T),\la\sigma_8\r(T)$ are determined as the minima of an
effective
potential $\hat U_{eff}(\sigma_0,\sigma_8)$. The effective potential is
calculated
as a constrained free energy density, i.e. the free energy density under the
constraint that the average fields $\frac{1}{\beta V}\int_0^\beta
d\tau\int d^3x\sigma_{0,8}(\vec x,\tau)$ take prescribed values
$\bar\sigma_0$ and $\bar\sigma_8$, respectively, while the same averages over
$\sigma_\ell,\ell=1,...,7$, and $\pi_\ell,\ell=0,...,8$ should vanish. The
physical
meaning is obvious. If one chooses $\bar\sigma_0,\bar\sigma_8\not=0$ in the
high-temperature chiral symmetric phase, the corresponding free energy is
certainly
not minimal for such a choice. We consider spontaneous symmetry breaking in two
directions. Accordingly, we introduce two background fields
$\bar\sigma_0,\bar\sigma_8$
\ba \label{ba}
\sigma_0&=&\bar\sigma_0+\sigma'_0\nonumber\\
\sigma_8&=&\bar\sigma_8+\sigma'_8, \ea
where $\sigma'_0,\sigma'_8$ denote the fluctuations around the background.
Otherwise, $\sigma_\ell=\sigma'_\ell$ for $\ell=1,...,7$ and
$\pi_\ell=\pi'_\ell$
for $\ell=0,...,8$. As a common notation for $\sigma'_\ell,\pi'_\ell$ we use
$M'_\ell=\sigma'_\ell+i\pi'_\ell$. The actual minima of $\hat
U_{eff}(\sigma_0,\sigma_8)$ (with
$\sigma_0=\bar\sigma_0,\sigma_8=\bar\sigma_8$) are
denoted as $\la\sigma_0\r$ and $\la\sigma_8\r$.

The Lagrangian is expanded in powers of $M'_\ell$. The linear term in $M'_\ell$
vanishes due to the $\delta$-constraint in the constrained free energy density.
The
quadratic term defines the masses $m^2_Q$  of the meson multiplets
$\pi,K,\eta,\eta',\sigma_\pi,\sigma_K,\sigma_\eta,\sigma_{\eta'}$, $Q=1,...,8$
labels
the multiplets. The explicit formulas were given in Eqs.~(\ref{aj}),(\ref{ak}).
The two
quartic terms are quadratized by introducing an auxiliary matrix field
$\sum(x)$
according to \cite{fre}
\ba \label{bb}
&&\exp\{-\beta[f_1({\rm Tr}M'M'^+)^2+f_2{\rm Tr}(M'M'^+)^2]\}=\nonumber \\ \\
&&=const\cdot\int^{c+i\infty}_{c-i\infty}{\cal D}\Sigma(x)\exp\{{\rm
Tr}\Sigma^2+2\ve{\rm Tr} (\Sigma M'M'^+)
+2\alpha{\rm Tr}(M'M'){\rm Tr}\Sigma\},\nonumber \ea
where $M'(x)$ is an $N\times N$-matrix field and
\begin{eqnarray*}
\ve^2&=&\beta f_2\\
2\ve\alpha+3\alpha^2&=&\beta f_1.
\end{eqnarray*}
Note that Eqs.~(\ref{bb}) are a sophisticated version of the simpler case,
where
$\phi$ is a scalar field
\be \label{bc}
const\cdot\exp\{-\alpha(\phi^2(x))^2\}=\int^{c+i\infty}_{c-i\infty}{\cal
D}\Sigma(x)
e^{\Sigma^2(x)-\Sigma(x)\phi^2(x)2\sqrt\alpha}. \ee
Formula (\ref{bc}) can be easily generalized to the case, where the l.h.s.
includes
also a cubic term $\phi^3(x)$, but we are not aware of an analogous
transformation
that leads to a tractable expression if the cubic term occurs in the form of a
determinant. This is the reason why we drop the cubic term in $M'_\ell$ in the
following procedure. The $(3\times3)$-matrix field $\Sigma(x)$ is replaced by
an
$SU(3)$-symmetric diagonal matrix $\Sigma={\rm diag}(sad,sad,sad)$. Thus the
matrix
of auxiliary fields is reduced to a single field variable $sad(x)$. The
quadratization of the quartic term ${\cal L}^{(4)}$ of the Lagrangian leads to
\ba \label{bd}
&&{\cal L}^{(4)}=f_1({\rm Tr}M'M'^+)^2+f_2{\rm Tr}(M'M'^+)^2\nonumber\\
\nonumber \\
&&\to{\cal
L}^{(4)\prime}=\frac{-3}{8(3f_1+f_2)}\left(\frac{1}{2}sad^2+\mu_0^2sad\right)+
\frac{1}{2}(s+\mu_0^2){\rm Tr}(M'M'^+).\ea

In the saddle-point approximation the path integral $\int{\cal D}\Sigma(x)$ is
dropped, the auxiliary field $sad(x)$ is replaced by $sad^*$, which maximizes
the
integrand. For an $O(N)$-model it is well known that this approximation
corresponds
to the leading order in a $1/N$-expansion \cite{pol}. In the special case of
$f_2=0=g$, the $SU(3)\times SU(3)$ sigma-model becomes an $O(18)$-model, which
is
invariant under $O(18)$-rotations. We have $N=18=2N_f^2$, where $N_f$ denotes
the
number of quark flavours, while $N$ labels the number of mesonic modes. Terms
of
$O(1/N)$ are dropped, as long as fluctuations in the auxiliary field are
neglected.
Therefore  we call our scheme a $large-N_f-approximation$.\\
The resulting one-loop contribution to the free energy density is given as
\be \label{be}
-\frac{1}{\beta V}\ln Z=\frac{1}{2\beta}\sum^8_{Q=1}g(Q)\sum_{n\in
Z}\int\frac{d^3K}
{(2\pi)^3}\ln(\beta^2(\omega^2_n+\omega^2_Q)), \ee
where
$g(\pi)=3,g(K)=4,g(\eta)=1,g(\eta')=1,g(\sigma_\pi)=3,g(\sigma_K)=4,g(\sigma_\eta)=1,
g(\sigma_{\eta'})=1$ are the multiplicity factors of the multiplets,
$\omega^2_n\equiv(2\pi n/T)^2$ and
\ba \label{bf}
\omega^2_Q&\equiv& K^2+X^2_Q\nonumber\\
X^2_Q&\equiv& sad+\mu^2_0+m^2_Q. \ea
Thus the one-loop contribution to the free energy takes a form, which is
familiar
from a free field theory. The only remnant of the interaction is hidden in the
effective mass square $X^2_Q$ via the auxiliary field variable $sad$. After the
sum
over the Matsubara frequencies is performed, the full expression for the
effective
potential is given as
\be \label{bg}
U_{eff}(\sigma_0,\sigma_8,sad)= U_{class}(\sigma_0,\sigma_8)+U_{saddle}(sad)+
U_0(\sigma_0,\sigma_8,sad)+U_{th}(\sigma_0,\sigma_8,sad), \ee
where $\sigma_0=\bar\sigma_0,\sigma_8=\bar\sigma_8$ and
\ba \label{bh}
&&U_{class}(\sigma_0,\sigma_8)=-\frac{1}{2}\mu^2_0(\sigma^2_0+\sigma^2_8)+\frac{g}
{3\sqrt3}(2\sigma^3_0-\sqrt2\sigma^3_8-3\sigma_0\sigma^2_8)\\
&&-\frac{2\sqrt2}{3}f_2\sigma_0\sigma^3_8+(f_1+\frac{f_2}{3})\sigma^4_0+(f_1+
\frac{f_2}{2})\sigma^4_8+2(f_1+f_2)\sigma^2_0\sigma^2_8-\ve_0\sigma_0-\ve_8\sigma_8
\nonumber\ea
is the classical part of the potential, which is independent of $sad$.
\be \label{bi}
U_{saddle}(sad)=-\frac{3}{8(3f_1+f_2)}\left(\frac{sad^2}{2}+\mu^2_0\ sad\right)
\ee
results from the transformation (\ref{bb}), cf. Eq.~(\ref{bd}). The one-loop
part
consists of the zero point energy $U_0$ and the thermal part $U_{th}$, given by
\be \label{bj}
U_0(\sigma_0,\sigma_8,sad)=\frac{1}{2}\sum^8_{Q=1}g(Q)\int^\Lambda
\frac{d^3K}{(2\pi)^3} \omega_Q, \ee
$U_0$ is divergent if the three-momentum cut-off $\Lambda$ is sent to infinity,
while
$U_{th}$ is convergent for $\Lambda\to\infty$ and vanishes for $T=0$
\be \label{bk}
U_{th}=\frac{1}{\beta}\sum^8_{Q=1}g(Q)\int\frac{d^3K}{(2\pi)^3}\ln(1-e^{-\beta
\omega_Q}). \ee
Our results have been obtained for the potential $\hat U_{eff}\equiv
U_{eff}-U_0$,
where the divergent zero-point energy has just been dropped. It is often argued
that
the omission of the zero-point energy is justified if one is finally interested
in
thermodynamic quantities, which are derived as derivatives of $\ln Z$ w.r.t.
$\beta$.
Usually, $U_0$ does not depend on $T$, and the splitting of the one-loop part
of
$U_{eff}$ according to $U_0$ and $U_{th}$ is a splitting in a $T$-independent
and
$T$-dependent part. Thus the contribution of $U_0$ should have no effect on the
thermodynamics. Strictly speaking we cannot use this argument, because $U_0$ of
Eq.~(\ref{bj}) has an implicit temperature dependence hidden in $X^2_Q$ via
$sad(T)$,
the temperature dependent saddle point variable, which is finally chosen such
that it
maximizes $\hat U_{eff}$ for $sad=sad^*$. This was one of the reasons why the
$U_0$-term was kept in \cite{met}. A renormalization prescription was imposed
such
that the strong (quartic) cut-off dependence of $U_{0}$ was weakened to a
$1/\Lambda^2$-dependence.
Further differences to \cite{met} are due to corrections of two errors in
\cite{met}.
After we had removed the programming error in $U_{class}$ of
\cite{met}, we found two branches in the free energy density $f$ (or the
pressure).
 The branches in $f$ of the high and low-temperature phases did not cross
at some temperature $T=T_c$, although the free energy should behave as a smooth
function in $T$ by general thermodynamic arguments. The reason was that the two
minima of the potential which exist below $T=200$ [MeV] were erroneously
associated  to
the true minima in the high- and low-temperature phases. It was not realized
that
for all temperatures up to $T\leq 200$ [MeV] one minimum is only a local one,
while
the other one stays the absolute minimum for all temperatures up to $T\geq200$
[MeV],
cf. also the remarks in section 6.

{\it Thermodynamic quantities}. Thermodynamic observables are derived from the
free
energy density $f=\lim_{V\to\infty}(-\frac{1}{\beta V}\ln Z)$, where $Z$ is
approximated as
\be \label{bm}
\hat Z(\la\sigma_0\r,\la\sigma_8\r)=e^{-\beta V\hat U_{eff}(\la\sigma_0\r(T),
\la\sigma_8\r
(T);sad^*(T))}. \ee
We have explicitly indicated that $\hat U_{eff}$ should be taken at the saddle
point
value $sad^*(T)$, which extremizes $\hat U_{eff}$. The expression for the
energy
density $\ve$, entropy density $s$, and the pressure $p$ are derived in the
standard
way. We have
\ba
p&=&-\hat U_{eff}, \label{bn}\\
s&=&-\left(\frac{\partial f}{\partial T}\right)=(p+\ve)/T, \label{bo}\\
\ve&=&-\frac{\partial \ln\hat
Z}{\partial\beta}=\sum^8_{Q=1}g(Q)\frac{T^4}{2\pi^2}
\int^\infty_0d\alpha\left(\frac{\alpha^2\sqrt{\alpha^2+y^2_Q}}{e^{\sqrt{\alpha^2
+y^2_Q}}-1}\right),\nonumber\\
\alpha&\equiv&K/T,\
y_Q\equiv\left(\sqrt{sad^*+\mu^2_0+m^2_Q(\la\sigma_0\r,
\la\sigma_8\r)}\right)/T. \label{bp}
\ea

{\it Evaluation of $\hat U_{eff}$}. Next let us discuss the evaluation of $\hat
U_{eff}$ in some more detail. The expression for $\hat U_{eff}$ is familiar
from
the expression for a free
field theory. In a high-temperature expansion it reads
\be \label{bq}
\hat
U_{eff}=U_{class}+U_{saddle}-\frac{T^4}{2\pi^2}\sum^8_{Q=1}g(Q)\left\{\frac{\pi^4}
{45}-\frac{\pi^2}{12}y^2_Q+O(y^3_Q)\right\} \ee
with $y_Q=\frac{X_Q}{T}$.

At high temperature the $SU(3)\times SU(3)$ linear sigma-model certainly fails
to
describe the quark-gluon plasma phase. The critical temperature falls neither
in a
high nor in a low-temperature region, the expansion parameter is of $O(1)$ or
larger.
Nevertheless, we discuss the high-temperature expansion, which has been
frequently
applied in calculations of a thermodynamic potential. It has the advantage that
imaginary parts of $\hat U_{eff}$ are absent to leading order in $X_Q/T$, cf.
e.g.
\cite{dol}. We have performed a high-temperature expansion for two sets of mass
parameters, the chiral limit and the realistic mass point, see sections 4 and
7. This
way we got some qualitative insight in the phase structure before we could
tackle the
problems in a fully numerical evaluation of $\hat U_{eff}$. Analytic
expressions for
a low-temperature expansion of $\hat U_{eff}$ are known as well \cite{hab}. We
have
used them only in intermediate steps to check the numerics. Thus we turn now to
the
fully numerical evaluation of $\hat U_{eff}$.

At a first glance the numerical evaluation of $\hat U_{eff}$ looks rather
straightforward. For each pair of $(\sigma_0,\sigma_8)$ we have to find $sad^*$
that
maximizes $\hat U_{eff}$. The condensates $\la\sigma_0\r,\la\sigma_8\r$ are
then
determined as minima of $\hat
U_{eff}(\sigma_0,\sigma_8;sad^*(\sigma_0,\sigma_8))$.
It is well known \cite{dol} that the arguments $(K^2+X^2_Q)$ of the logarithm
in
$\hat U_{eff}$ can become negative and lead to imaginary parts in $\hat
U_{eff}$,
which we have mentioned above. The original hope was that the positive
contribution
of the auxiliary field $sad$ to the masses $m_Q$ helps in avoiding imaginary
parts of
$\hat U_{eff}$.
\begin{figure}[tbh]
\vspace{13.0cm}
\caption{Saddle point $sad^*$ as a function of temperature $T$}
\end{figure}
Actually no imaginary parts have been found in \cite{fre}. As
can be seen from Fig.~2 the contribution of $sad^*$ to $X^2_Q$ increases with
$T$, it
is positive for $T\geq 116$~[MeV], but in our case the positive contribution is
not
sufficient to avoid negative $X^2_Q$ completely, i.e. for all
$\sigma_0,\sigma_8$ and
$T$. It has turned out that the actual minima $\la\sigma_0\r,\la\sigma_8\r$ lie
always
in the ``allowed'' region of real valued $\hat U_{eff}$ or at least at the
boundary
of this region, but on the way of searching the maximum in $sad^*$ and the
minima in
$\la\sigma_0\r,\la\sigma_8\r$ the routines encounter negative mass squares
indispensably. Therefore $\hat U_{eff}$ is analytically continued in the
following
way. The integral in $U_{th}$ of $\hat U_{eff}$ is of the type
\be \label{br}
I(z)=\int^\infty_0 dx\{x^2\ln[1-\exp[-(x^2+z^2)^{1/2}]\}
=-z^2\sum^\infty_{n=1}\frac{1}{n^2}K_2(nz), \ee
where $K_2$ is a modified Bessel function. The analytic continuation of
$K_m(nz)$ to
complex values of $nz$ is given as
\be \label{bs}
K_m(nz)=i^{m+1}\left(\frac{\pi}{2}\right)[J_m(inz)+iY_m(inz)], \ee
where $J_m$ are Bessel functions of the first kind and $Y_m$ are Weber
functions
\cite{abr}. This leads to the following form of $\hat U_{eff}$
\ba \label{bt}
\hat
U_{eff}&=&U_{class}+U_{saddle}+\frac{T^4}{2\pi^2}\sum^8_{Q=1}g(Q)\left\{\frac{-
X^2_Q}{T^2}\sum^\infty_{n=1}\frac{1}{n^2}K_2\left(\frac{n}{T}X_Q\right)\right\}
\ea
for $X^2_Q\geq0$. For $X^2_Q<0$ and $X_Q = \pm i \sqrt{|X_Q^2|}$
\ba
\hat U_{eff}&=&U_{class}+U_{saddle}+{\rm Re}\ U_{th}+i\ {\rm Im}\ U_{th}
\nonumber
\ea
with
\ba \label{bu}
{\rm Re}\ U_{th}&=&\frac{T^4}{2\pi^2}\sum^8_{Q=1}g(Q)\left\{\frac{\pi}{2}
\frac{|X^2_Q|}{T^2}\sum^\infty_{n=1}
\frac{1}{n^2}Y_2\left(\frac{n}{T}|X^2_Q|^{1/2}\right)
\right\},\nonumber\\
{\rm Im}\
U_{th}&=&\frac{T^4}{2\pi^2}\sum^\infty_{Q=1}g(Q)\left\{\pm\frac{\pi}{2}
\frac{|X^2_Q|}{T^2}\sum^\infty_{n=1}\frac{1}{n^2}J_2\left(\frac{n}{T}|X^2_Q|^{1/2}\right)
\right\}. \ea
In the actual calculation we have cut the infinite sums in
Eq.~(\ref{bt}),(\ref{bu})
after 25 terms.\\
The condensate values $\la\sigma_0\r(T),\la\sigma_8\r(T)$ are determined as
minima of
$\hat U_{eff}$ for $X^2_Q\geq0$ and of ${\rm Re}\ \hat U_{eff}$ for $X^2_Q<0$.

{\it Error analysis}. There are mainly two sources for errors. The first one is
the
determination of the saddle point from the condition $\partial\hat
U_{eff}/\partial
sad=0$. In searching $sad^*$ the routines encounter negative values of $X^2_Q$
in
intermediate steps and find $sad^*$ from derivatives of Weber functions
$Y_2(\frac{n}{T}|X^2_Q|^{1/2})$, see Eq.~(\ref{bu}). The Weber functions are
taken
from representations in \cite{pre} which cover the range from small to large
arguments.
For arguments
$\frac{n}{T}|X^2_Q|^{1/2}\geq 4.8 (9.7), Y_2(\frac{n}{T}|X^2_Q|^{1/2})\leq
6.0\cdot 10^{-3} (2.9\cdot 10^{-5})$. Due to the large variation of the
effective
masses the arguments become easily $\geq 100$.
Thus for small temperatures, large effective masses and/or large values of $n$
the contributions to the sum over $n$ are so small that they reach the order of
the
numerical accuracy. The location of the saddle point becomes inaccurate under
these
conditions. Similarly, for small arguments (i.e. high temperatures, small
values of
$X_Q$) the approximations of the Weber functions in the vicinity of their
singularity
at vanishing arguments become less reliable \cite{pre}. This explains why the
error in $sad^*$ increases with temperature, if simultaneously $|X^2_Q|^{1/2}$
becomes smaller.\\
The error $\Delta sad^*$ in the maximum $sad^*$ of $\hat U_{eff}$ is estimated
as the
difference between the zeroes of $\partial\hat U_{eff}/\partial sad=0$ if
$sad^*$ is
extrapolated from the right or approached from the left in the
following way. On the left hand side of the
maximum, ${\rm Re}\ \hat U_{eff}$ is represented by the series of Weber
functions, it
is oscillating and $\partial \hat U_{eff}/\partial sad>0$. The zero of the
oscillating derivative is actually used in the program as value for $sad^*$.
The
oscillations extend to some region (of temperature dependent size) on the right
hand
side of $\partial \hat U_{eff}/\partial sad=0$, where the derivative is
negative. For
further increasing values of $sad$ the effective potential becomes real and
behaves
approximately quadratically in $sad$, such that $\partial\hat U_{eff}/\partial
sad$
is linear in $sad$. We extrapolate the linear part of the derivative to zero
and use
the extrapolated zero as second value for $sad^*$. The difference between both
values
for $sad^*$ gives an estimate for the ambiguity in finding the maximum of $\hat
U_{eff}$ from the parabolic or the oscillating behaviour. It turns out that for
low
temperatures $\Delta sad^*\sim0$, while it increases with temperature to
$\Delta
sad^*\sim\pm0.006$~[GeV]$^2$ in the transition region in case of the
chiral limit, cf.~Fig.~2. \\
The uncertainty in the
determination in $sad^*$ leads to errors in the meson condensates, the
effective
masses $X_Q$, and the thermodynamic quantities $\ve,p$, and $s$. The errors
$\Delta\la\sigma_{0,8}\r$ which are induced in $\la\sigma_0\r,\la\sigma_8\r$
due to the
uncertainty in the precise value of $sad^*$ are estimated from
\be \label{bv}
\Delta\la\sigma_{0,8}\r=\left.\frac{-\partial^2\hat U_{eff}/\partial
sad^*\partial\sigma_0}
{\partial^2U_{eff}/\partial^2\sigma^2_0}\right|_{\la\sigma_0\r,\la\sigma_8\r,sad^*}
\cdot\Delta sad^*, \ee
since $\la\sigma_{0,8}\r(sad^*)$ are implicitly determined via
$\partial\hat U_{eff}/ \partial\sigma_{0,8}=0$. In case of realistic
masses this leads to
$\Delta\la\sigma_{0,8}\r\leq10^{-7}$ below $T\sim50$~MeV, $\leq10^{-4}$ below
$T\sim125$~MeV, $\leq10^{-3}$ for $T\leq165$~MeV and of the order of
$10^{-3}$~[MeV] in the transition region. After the crossover ($T\geq197$~MeV)
it stays $\sim O(10^{-4})$.

An independent second source for errors in the condensates
$\la\sigma_0\r,\la\sigma_8\r$ is the flat shape of the effective potential in
the
transition region, in particular for external field strengths which only admit
a
rather weak first order phase transition. We estimate $\Delta \la \sigma_0 \r (
\simeq
\Delta \la \sigma_8\r )$ from plots of $\hat U_{eff}(\sigma_0)$ in the
$SU(3)$-symmetric
case. We find $\Delta\la\sigma_0\r \simeq \pm 1.3$~[MeV] in the chiral limit
and
$ \simeq \pm 4.3$~[MeV] for $\epsilon_0 = 2\cdot 10^{-4}$ or
$2.5 \cdot 10^{-4}$~[GeV]$^3$, cf.~Fig.~6 of section 5.\\
In the $SU(3)$-symmetric case
$\la\sigma_8\r$ should identically vanish for all temperatures. The eight
Goldstone
bosons are identical in mass. If the $SU(3)$-symmetric case is calculated as a
special
case ($\ve_0\not=0=\ve_8$) of our general framework ($\ve_0\not=0\not=\ve_8),\
\la\sigma_8\r=0$ should come out automatically. In fact it vanishes within the
numerical accuracy for temperatures $T\leq 50$~[MeV] for $\epsilon_0 = 2.5\cdot
10^{-4}$~
[GeV]$^3$. In the transition region
$\la\sigma_8\r$ strongly fluctuates around zero. The maximal fluctuation
varies between $0.2$ and $3.0$~[MeV] depending on $\epsilon_0$ and the
strength of the first order transition. The values are of the same order of
magnitude as $\Delta\la\sigma_{0,8}\r$ due to the flat shape of the effective
potential.\\
When the meson condensates are converted to the quark condensates, additional
errors
enter due to the current quark masses. We use $\Delta m_{u,d}=\pm1.45$~[MeV]
\cite{nar}.\\
Finally, the errors in the entropy and energy densities $s$ and $\ve$ can be
traced
back to the effective masses $X_Q(\la\sigma_0\r,\la\sigma_8\r,sad^*)$.
The resulting errors in $s/T^3$ in the crossover region are indicated in
Fig.~10 of section 6, where $s/T^3$ is
shown for realistic meson masses.

 From very different approximation schemes it is known that the region of a
phase
transition is particularly difficult to handle. We consider our difficulties
in localizing the
saddle point and calculating the condensates in the transition/crossover region
just as
a special manifestation of that.

\section{The chiral limit}

 From the renormalization group analysis of Pisarski and Wilczek \cite{pis} we
expect
a first order chiral transition in the chiral limit. As mass input we choose
$m^2_\pi=m^2_K=m^2_\eta=0$ for the pseudoscalar Goldstone bosons,
$m_{\eta'}=850$~[MeV], $m_{\sigma_{\eta'}}=800$~[MeV], $f_\pi=94$~[MeV]. The
mass of the
sigma meson should be treated as a parameter due to the uncertainty in the
experimental identification. For the final parametrization we choose
$m_{\sigma_\eta}=600$~[MeV].

The tree level parametrization of the Lagrangian (\ref{aa}) in the
limit of vanishing $\ve_0$ and $\ve_8$ is then given by
\ba \label{ca}
\mu^2_0&=&5.96\cdot 10^{-2}\ {\rm [GeV]}^2\nonumber\\
f_1&=&4.17\nonumber\\
f_2&=&4.48\nonumber\\
g&=&-1.81 {\rm [GeV]}. \ea
\begin{figure}[tbh]
\vspace{13.0cm}
\caption{Normalized light quark condensate $\la\bar qq\r_T / \la\bar qq\r_0$ as
a
function of
temperature $T$ in the
high-temperature expansion (solid curve) and the numerical evaluation (dashed
curve)
of the large $N_f$-expansion. The condensate drops to zero at $T_c\sim92$~[MeV]
and
177~[MeV], respectively.} \end{figure}
The high-temperature expansion gives a 1st order
transition at $T_c=92\pm1$~[MeV], while the fully numerical evaluation leads to
$T_c=177\pm1$~[MeV], cf. Fig.~3. This is in agreement with the general
observation that
the high-$T$-expansion gives qualitatively correct results when it is
extrapolated
beyond its validity range, quantitatively it fails in precise predictions.
The $T_c$-value of the numerical calculation supports the estimate for the
$N_f$-dependence of $T_c$, which has been derived by Cleymans, Koci\'c and
Scadron~\cite{cley} from a pion gas model without interactions as
\be\label{xa}
T_c\sim 2\cdot f_\pi\cdot\sqrt{3N_f/(N_f^2-1)}\sim 200{\rm [MeV]}.
\ee

The effective potential $\hat U_{eff}$ is plagued with imaginary parts for all
temperatures we have investigated, i.e. between 0 and 250~[MeV]. When we
evaluate ${\rm
Re}\ \hat U_{eff}$ according to Eq.~(\ref{bu}), we observe the same effect for
higher
values of the sigma-mass as it has been noticed by Goldberg \cite{gol}.
Goldberg has
calculated the effective potential of the $SU(3)\times SU(3)$ sigma-model in a
mean-field approximation in the chiral limit. The oscillations in the potential
became
stronger for larger values of the $\sigma$-meson mass, which was used as input.
In our
case ${\rm Re}\ \hat U_{eff}(\sigma_0)$ strongly fluctuates around the expected
parabola if $m_\sigma\geq1$~[GeV]. This explains our final parameter choice of
$m_{\sigma_\eta}=600$~[MeV]. The approximation of ${\rm Re}\ \hat U_{eff}$
according to
Eq.~(\ref{bu}) by a finite series of Weber functions $(n\leq25)$ loses its
validity if
the argument of $Y_2$ becomes too large due to $m^2_\sigma$.\\
For the barrier height
between the coexisting minima of the potential we find $0.14\cdot
10^{-3}$~[GeV/fm$^3$]. The barrier height may be regarded as  a measure for the
strength of the transition. It determines the tunneling rate between coexisting
phases
in the transition region. The  barrier height is clearly smaller than the value
of
 Frei and Patk\'os \cite{fre}, who found
0.36~[GeV/fm$^3]$ in the same model, but without inclusion of the
$n\not=0$-Matsubara
frequencies. The higher barrier goes along with a large value for the interface
tension $\alpha$ between the coexisting chirally broken and symmetric phases at
$T_c$, $\alpha$ has been estimated as $[(40-50) {\rm MeV}]^3$~\cite{fre}. More
recent
lattice results indicate that such a large value is not likely for
QCD~\cite{kaj}.

Of particular interest in the
chiral limit is the temperature dependence of the effective masses
$X^2_Q=sad+\mu^2_0+m^2_Q,\ Q=1,...,8$. In the chiral limit the masses are free
of
contributions of the external fields. Thus the Goldstone theorem should be
obeyed by
the pseudoscalar meson octet.
\begin{figure}[tbh]
\vspace{13.0cm}
\caption{Effective masses $X_Q,\ Q=1,...,8$ as a function of $T$ in the chiral
limit.
In units of [MeV] we have at $T=0\ X_\pi=X_K=X_\eta=0,X_{\eta'}=850$,
$X_{\sigma_{\eta'}}=600$, $X_{\sigma_\pi}=X_{\sigma_K}$=$X_{\sigma_\eta}=800$.
The
masses are degenerate at $T_c\sim177\pm3$~[MeV].}
\end{figure}
A short glance at Fig.~4 shows
that the effective masses $X_Q=X_\pi=X_K= X_\eta$ do not stay massless for
$T>0$. They get an increasingly positive contribution with increasing $T$ due
to the
saddle point contribution $sad^*(T)$. If the physical meaning of $X_Q$
coincides to
leading order in $1/N_f$ with a dynamical mass (in particular if the dynamical
and
screening masses coincide to leading order in $1/N_f$ and both are equal to
$X_Q$),
the increase of $X_Q(T)$ with $T$ would violate Goldstone's theorem and should
be
considered as an artifact of our approximation. We still have to clarify this
point.

\section{The critical transition line}

The critical transition line in an $(m_\pi,m_K,\ldots)$-diagram consists of
pseudoscalar
meson masses for which the first order chiral transition becomes of  second
order and
turns into a crossover phenomenon for meson masses exceeding the critical
values. We
will determine  three such critical points.  The critical points  are
characterized
by their external field strengths $\ve_0,\ve_8$. The first point corresponds to
 an
$SU(3)$-symmetric case, where $\ve_8=0,\ m_u=m_d=m_s\not= 0$ and
$m_\pi=m_K=m_\eta\not= 0$. Since  $<\sigma_8>=0$  for all temperatures, the
numerics
considerably simplifies compared to the general case with
$\ve_0\not=0\not=\ve_8$.
For the second critical point we choose $\ve_0=-0.77\ \ve_8$. This ratio is
identical to $\ve_0/\ve_8$ for the mass point  with realistic meson masses,
where
$m_s/\hat m=18.2$, cf. section 2.  The third point is characterized by
$\ve_0=(2\alpha/\beta)\cdot \ve_8$. It is chosen such that $m_s=0,\
m_{u,d}\not=0$.
Before we present our results for the critical field strengths in the
large-$N_f$-expansion, we calculate $\ve_0^{crit},\ \ve_8^{crit}$ in a
mean-field
approximation.

\subsection{Critical meson masses in a mean-field approximation}

Recently, Gavin, Goksch and Pisarski~\cite{gav} have calculated a set of
critical
quark masses in the linear $SU(3)\times SU(3)$ sigma-model. The calculation has
been
performed in a mean-field approximation. Since we use a  different tree-level
parametrization of the sigma-model, we have reproduced the mean-field
calculation for
our parameter choice and summarize the main steps.

{\it The $SU(3)$-symmetric case.} Let us first consider the $SU(3)$-symmetric
case
with $\ve_8=0$. In mean-field we have to deal with the classical part of the
potential $U_{class}$, which follows from the Lagrangian (\ref{aa})  for a
constant
background field $\sigma_0=\bar\sigma_0\ (\bar\sigma_8=0$). We have
\be\label{cb}
U_{class}(\sigma_0)=-\frac{1}{2}\mu^2_0\sigma^2_0+\frac{2}{3\sqrt3}g\sigma^3_0+(f_1+
\frac{f_2}{3})\sigma^4_0-\ve_0\sigma_0.
\ee
This form is familiar from  Landau's free energy functional in terms of an
order
parameter field. One of its applications is a description of the phase
structure for
a liquid/gas transition. For $\ve_0=0$ the system has a first order transition
from
the liquid to the gas phase. The transition stays first order, until the
external
field (the pressure in case of a liquid/gas system) is increased to a critical
value
$\ve_0^{crit}$, where it becomes second order. For values of
$\ve_0>\ve_0^{crit}$, the
transition is washed out and turns into a crossover phenomenon.

In  mean-field the effect of a finite (and strictly speaking high) temperature
is
reduced to a renormalization of the mass parameter term, i.e. of the
coefficient of
the quadratic term in the Lagrangian.  Thus the finite temperature effects can
be
mimiced by tuning  $\mu^2_0$ while keeping the other couplings in the
Lagrangian
$(f_1,f_2,g)$  fixed.  The potential can be Taylor expanded around its true
minimum
$\sigma_0^{min}$ (which is different from zero in the symmetry broken phase);
in
particular it can be expanded around the ``critical'' $\sigma^{min}_0\equiv
\sigma_0^{crit}$ for the critical field strength $\ve_0^{crit}$. In the
``critical''
case  $U_{class}$ starts with a term quartic in $(\sigma_0-\sigma_0^{crit})$.
{}From
the vanishing of the first three coefficients we obtain the critical parameters
as
follows.\\
Since the first order transition just disappears for $\ve_0^{crit}$,  the third
derivative of $U_{class}$ w.r.t. $\sigma_0$ should vanish at
$\sigma_0=\sigma_0^{crit}$ leading to
\be\label{cc}
\sigma_0^{crit}=\frac{-g}{\sqrt3\cdot 6(f_1+f_2/3)}=3.1\cdot 10^{-2} {\rm
[GeV]}.
\ee
The second derivative $\partial^2 U_{class}/\partial\sigma^2_0$ at
$\sigma_0^{crit}$ is
the coefficient of the quadratic  fluctuations $(\sigma_0-\sigma_0^{crit})^2$
around
$\sigma_0^{crit}$, thus it has the meaning of $m^2_{\sigma_{\eta'}}$. This is
the
critical mass, which goes to zero when the second order transition is
approached.  It
is easily checked  that the other meson masses remain finite for the same set
of
critical parameters. The vanishing of
$m^2_{\sigma_{\eta'}}$ at
criticality or $\partial^2
U_{class}/\partial\sigma^2_0$ at $\sigma_0^{crit}$ implies for the critical
value of $\mu^2_0$
\be\label{cd}
\mu^{2\ crit}_0=\frac{-g^2}{9\cdot(f_1+f_2/3)}=-6.44\cdot 10^{-2}\ {\rm
[GeV]}^2.
\ee
\noindent Finally the extremum condition for $\sigma_0^{crit}$ determines
$\ve_0^{crit}$ as
\be\label{ce}
\ve_0^{crit}=-\frac{1}{27\cdot 6\cdot \sqrt3}\frac{g^3}{(f_1+f_2/3)^2}=6.6\cdot
10^{-4}\ {\rm [GeV]}^3.
\ee

The value for $\ve_0^{crit}=6.6\cdot 10^{-4}\ {\rm [GeV]}^3\ (\ve_8^{crit}=0)$
leads to a pseudoscalar octet mass of $m_\pi=m_K=m_\eta=146$ [MeV] and to a
current
quark mass of $m_u=m_d=m_s=1.9$ [MeV]. One should keep in mind that the tree
level
parametrization of the  $SU(3)\times SU(3)$ sigma-model is arbitrary to some
extent.
If we would  choose $g=-1.39\ {\rm [GeV]}, \ f_1=5.3,\ f_2=0.93$
with  $m_{\sigma_{\eta'}}=600\ {\rm [MeV]}$ (the values which have
been used in~\cite{mey} for the tree level parametrization), $\ve_0^{crit}$
turns out
as $3\cdot 10^{-4}\ {\rm [GeV]}^3$ or $m_{u,d,s} \simeq 0.9$~[MeV] leading
to $\la m_{ps} \r = 115$~[MeV]. For the same parameter choice, but
$f_1 = 2.35$ and
$m_{\sigma_{\eta}}=950\ {\rm [MeV]}, \ve_0^{crit}$ comes out as $1.3\cdot
10^{-3}\
{\rm [GeV]}^3$. The same tendency has been observed in~\cite{gav}. An increase
of the
input mass $m_{\sigma_\eta}$ shifts the critical field strength to larger
values
(reducing the discrepancy to the lattice result).
 The parameter choice which has been used in~\cite{met} leads
to $\ve_0^{crit}=4\cdot 10^{-4}\ {\rm [GeV]}^3$. Gavin, Goksch and
Pisarski~\cite{gav} obtain for the corresponding critical field
strength in the $SU(3)$-symmetric case $h_0^{crit}=1.6\cdot 10^{-4}\ {\rm
[GeV]}^3$.
Due to their different parametrization it is not identical with ours, but of
the same
order of magnitude.
Thus we estimate the error in the
mean-field  calculation as $\Delta\epsilon_0^{crit} = \pm 5\cdot 10^{-4}
\ {\rm [GeV]}^3$. The induced error in the pseudoscalar meson masses comes out
as
large as $\approx 127$~[MeV], in the current light quark mass it is $\leq
1.4$~[MeV].
\\

{\it The general case of} $\ve_0\not= 0,\ \ve_8\not= 0$. For non-vanishing
$\ve_0$
and $\ve_8$ $U_{class}$ is given by Eq.~(\ref{bh}). Again we have to determine
the
critical parameter $\mu^2_0$ to simulate a finite temperature, the critical
minima
values $\sigma_0^{crit},\ \sigma_8^{crit}$ of $U_{class}$ and the critical
field
strengths $\ve_0^{crit},\ \ve_8^{crit}$. We use the following conditions
\ba
\partial U_{class}(\sigma_0,\sigma_8)/\partial\sigma_0\Bigm\vert_{crit}&=&0
\label{cf}\\
\partial U_{class}(\sigma_0,\sigma_8)/\partial\sigma_8\Bigm\vert_{crit}&=&0
\label{cg}\\
m^2_{\sigma_{\eta'}}\Bigm\vert_{crit}&=&0\label{ch}\\
\ve_0^{crit}/\ve_8^{crit}+0.77\Bigm\vert_{crit}&=&0\label{ci}\\
\partial^3 U_{class}(\sigma_0,\sigma_8)/\partial
r^3\vert_{crit}&=&0\;.\label{cj}
\ea
\noindent The five conditions are postulated at criticality (abbreviated as
$\Bigm\vert_{crit}$),
i.e. for the set of critical parameters $\sigma_0^{crit},\ \sigma_8^{crit},\
\ve_0^{crit},\  \ve_8^{crit},\ \mu^{2\ crit}_0$. The first three equations are
obvious generalizations of the $SU(3)$-symmetric case. Eq.~(\ref{ci}) is just
one
possible choice  saying that we keep the mass splitting of realistic
(pseudo)scalar
meson masses fixed in the tuning  to the critical  transition line.
Eq.~(\ref{cj})
generalizes $\partial^3
U_{class}(\sigma_0)/\partial\sigma^3_0\bigm\vert_{crit}=0$
of the one-dimensional case. More precisely, it postulates that the directional
derivative in the radial direction $r$ in $(\sigma_0,\sigma_8)$-space (i.e.
perpendicular to the direction of $m^2_{\sigma_{\eta'}}=0$) should vanish to
exclude
the occurrence of a first order transition.
Eqs.~(\ref{cf}-\ref{cg}) lead to
\ba\label{ck}
\ve_0^{crit}&=&-\mu^2_0\sigma_0+\frac{g}{\sqrt3}(2\sigma_0-\sigma_8)-\frac{2}{3}
\sqrt2
f_2\sigma^3_8+4(f_1+f_2/3)\sigma_0+4(f_1+f_2)\sigma_0\sigma_8^2\nonumber\\
\ve_8^{crit}&=&-\mu^2_0\sigma_8-\sqrt{\frac{2}{3}}\cdot g(\sigma^2_8+\sqrt2
\sigma_0\sigma_8)-2\sqrt2 f_2\sigma_0\sigma^2_8+\nonumber\\
&&+4(f_1+f_2/2)\sigma^3_8+4(f_1+f_2)\sigma^2_0\sigma_8,
\ea
\noindent where finally $\mu^2_0=\mu^{2\ crit}_0,\ \sigma_0=\sigma_0^{crit},\
\sigma_8=\sigma_8^{crit}$.\\
If we introduce the auxiliary quantities $r_0,r_8$ according to
\ba\label{cl}
r_0&\equiv&(12
f_1+4f_2)\sigma^2_0+4(f_1+f_2)\sigma^2_8+4g\sigma_0/\sqrt3\nonumber\\
r_8&\equiv&4(f_1+f_2)\sigma^2_0+6(2f_1+f_2)\sigma^2_8-4\sqrt2
f_2\sigma_0\sigma_8-\frac{2g}{\sqrt3}\sigma_0-2\sqrt{\frac{2}{3}} g\sigma_8
\ea
\noindent and use $m^2_{\sigma_{08}}$ of Eq.~(\ref{ak}), Eq.~(\ref{ch}) implies
\be\label{cm}
\mu^{2\ crit}_0 =\frac{1}{2}\left[
r_0+r_8-\sqrt{(r_0-r_8)^2+4(m^2_{\sigma_{08}})^2}\right],
\ee
\noindent where in the end  $\sigma_0=\sigma_0^{crit},\
\sigma_8=\sigma_8^{crit}$ in
Eqs.~(\ref{cl}).

So far we have $\ve_{0,8}^{crit}=\ve_{0,8}^{crit}(f_1,f_2,g,\sigma_0,\sigma_8)$
and
$\mu_0^{2\ crit}=\mu_0^{2\ crit}\ (f_1,f_2,g,\sigma_0,\sigma_8)$.  Finally the
zeroes
of Eqs.~(\ref{ci}-\ref{cj})  determine $\sigma_0^{crit},\ \sigma_8^{crit}$.
These
equations are solved numerically. We find
\be\label{cn}
\sigma_0^{crit}=3.6\cdot 10^{-2}{\rm [GeV]},\ \sigma_8^{crit}=-5.5\cdot
10^{-3}{\rm
[GeV]}. \ee
\noindent Using these values, Eqs. (\ref{cm}), (\ref{ck}) and (\ref{cl}) give
\ba
\mu^{2\ crit}_0&=&-6.5\cdot 10^{-2}{\rm [GeV]}^2,\label{co}\\
\ve_0^{crit}&=&7.7\cdot 10^{-4}{\rm [GeV]}^3,\ \ \ \ \ve_8^{crit}=-1.0\cdot
10^{-3}{\rm
[GeV]}^3\label{cp} \ea
\noindent
or $m_s = 6$~[MeV], $\hat m = 0.3$~[MeV], while the average pseudoscalar meson
mass
$\la m_{ps} \r$ = $126.4$~[MeV].
The values for $\ve^{crit}_{0,8}$ are compatible with the results of Gavin,
Goksch and Pisarski~\cite{gav}, who find for the critical field strengths
$h_0^{crit}=(62\ {\rm [MeV]})^3,\ h_8^{crit}=(60.4\ {\rm [MeV]})^3$, if one
keeps in
mind  the different tree level parametrization. For example $m_s/m_{u,d}=32$
in~\cite{gav}, while $m_s/m_{u,d}=18.2$ in our case.

{\it The case of $m_s=0$ or $\ve_0/\ve_8=2\alpha/\beta$.} If we replace $0.77$
in Eq.~(\ref{ci}) by $-2\alpha/\beta=-1.31$ corresponding to  $m_s=0,\
m_{u,d}\not=0$,
we obtain from Eqs.~(\ref{cf})-(\ref{cj})
\ba
\sigma_0^{crit}&=&3.4\cdot 10^{-2}{\rm [GeV]},\ \ \ \ \sigma_8^{crit}=2.9\cdot
10^{-3}{\rm
[GeV]}\label{cq}\\
\mu^{2\ crit}_0&=&-6.4\cdot 10^{-2}{\rm [GeV]}^2\label{cr}\\
\ve_0^{crit}&=&6.7\cdot 10^{-4}{\rm [GeV]}^3,\ \ \ \ \ve_8^{crit}=5.15\cdot
10^{-4}{\rm
[GeV]}^3\label{cs} \ea
\noindent or $\hat m = 2.9$~[MeV] and $\la m_{ps} \r = 137.2$~[MeV].
Note that the condition $m_s=0$ implies the same sign for $\ve_0^{crit}$ and
$\ve_8^{crit}$ due to the same sign for the constants $\alpha$ and $\beta$.
Hence the
``critical'' condensates $\sigma^{crit}_{0,8}$ come out with equal sign  in
contrast
to the realistic mass case.\\
In the following we will compare the mean-field values
for $\ve^{crit}_{0,8}$ for  Eqs.~(\ref{ce}), (\ref{ck}), (\ref{cs}) with the
large-$N_f$-results.

\subsection{Critical meson masses in the large-$N_f$-expansion}

{\it The $SU(3)$-symmetric case.}  When $\ve_8=0$ and $\ve_0$ is slowly
increased from
0 to $\sim 2.5\cdot 10^{-4}$ [GeV]$^3$, we observe a weakening of the first
order
transition as  is seen in Fig. 5. For the critical field strength we find
\be\label{ct}
\ve_0^{crit}\leq (3\pm 0.5)\cdot 10^{-4}{\rm [GeV]}^3.
\ee
\begin{figure}[tbh]
\vspace{13.0cm}
\caption{The light quark condensate normalized to its value at zero temperature
$\la\bar q q\r_T/\la\bar q q\r_0$ as a function of $T$ in the $SU(3)$-symmetric
case.
The weakening of the first order transition is obvious, when $\ve_0$~[GeV]$^3$
is
varied between $\ve_0=0\ (\times), 2\cdot10^{-4}\ (\Diamond), 2.5\cdot10^{-4}\
(+)$
and $6.6\cdot10^{-4}\ (\Box)$.}
\end{figure}
The value for $\ve_0^{crit}$ induces a pseudoscalar meson mass of 51 [MeV] or
quark
masses of $m^{crit}_{u,d}=m_s^{crit}=0.9 \pm 0.14$ [MeV], such that
\be\label{cu}
m^{crit}_{u,d,s}/m_{u,d,s}\sim 0.08 \pm 0.01 .
\ee
Within the errors
the result
for $\ve^{crit}_0$
is of the same order of magnitude as the mean-field value. The uncertainty
in our result is $\Delta \ve_0^{crit}\leq0.5\cdot10^{-4}$ [GeV]$^3$.
For $\ve_0=2.5\cdot 10^{-4}$ [GeV]$^3$ the transition could be clearly
identified as
first order from the $\la \bar qq\r_T / \la \bar qq\r_0 (T)$-curves.
For  $\ve_0=3\cdot 10^{-4}$ [GeV] it
is a crossover. We could have further improved the accuracy of $\ve_0^{crit}$
by
measuring data for $\la \bar q q\r_T (T)$ in the intermediate $\ve_0$-range. On
the
other hand such an improvement is anyway limited by the well-known fact that it
is in
general hard to disentangle a very weak first order transition from a rapid
crossover
phenomenon.

The weakening of the first order transition is also revealed in the barrier
height of
the effective potential between the coexisting minima at $T_c$. The barrier
decreases
from $1.4\cdot10^{-4}$~[GeV/fm$^3$] in the chiral limit to
$2.1\cdot10^{-6}$~[GeV/fm$^3$] for $\ve_0=2.0\cdot10^{-4}~{\rm [GeV]}^3,\
\ve_8=0\ {\rm
[GeV]}^3$, the largest value for which a first
 order transition
could be identified from the shape of the effective potential, cf. Fig. 6. A
comparison between Figs.~5 and 6 shows the ambiguity in identifying a very weak
1st
order transition. Fig.~5 suggests a weak 1st order transition for
$\ve_0=2.5\cdot10^{-4}$~[GeV$]^3$ with $T_c\sim181$~[MeV], but no barrier is
visible
between the two coexisting condensate values at the same temperature and the
same
$\ve_0$-value in Fig.~6.
Accordingly a precise determination of $T_c$ is hard in case of a weak first
order
transition such that we estimate for the error in $T_c$
$\Delta T_c = \pm 2$ [MeV].
Fig.~6 also admits an estimate of the error in finding
$\la\sigma_{0,8}\r$ in the transition/crossover region if the effective
potential is
very flat, which we have mentioned in section 3. In the chiral limit we have
$\Delta\la\sigma_0\r\sim1.3\cdot10^{-3}$~[GeV], for $\ve_0=2\cdot10^{-4}$ and
$2.5\cdot10^{-4}$ the error is estimated as $4.3\cdot10^{-3}$~[GeV]. Later we
assume
$\Delta\la\sigma_0\r\sim\Delta\la\sigma_8\r$.
\begin{figure}[tbh]
\vspace{13.0cm}
\caption{Decrease of the  barrier height of the effective potential $\hat
U_{eff}$
for $\ve_0=0$ at $T\sim176$~[MeV] (solid curve),
$\ve_0=2\cdot10^{-4}$~[GeV]$^3$ at
$T\sim180$~[MeV]$^3$ (dashed curve), $\ve_0=2.5\cdot10^{-4}$~[GeV]$^3$
for $T\sim181$~[MeV]  (determined from the corresponding
condensate curve of Fig.~5) (dotted curve) in the $SU(3)$-symmetric case.}
\end{figure}


{\it $T_c(m)$-dependence in the $SU(3)$-symmetric case.} For simplicity we
restrict
the study of the mass dependence of $T_c$ in the first order transition region
to the
$SU(3)$-symmetric case.  In Fig. 7 we see an increase of $T_c$ with the
current quark mass $\hat m$~[MeV].
\begin{figure}[htb]
\vspace{13.0cm}
\caption{The critical temperature $T_c$ as a function of the
average light quark mass $\hat m$ [MeV]
in the $SU(3)$-symmetric case.  For further explanations see
the text.}
\end{figure}
For realistic mass values of
$m_\pi=129.3$ [MeV], $m_K=490.7$ [MeV], $m_\eta=544.7$ [MeV] (not depicted in
Fig.7),
the rapid crossover sets
in at $T\sim 181.5$ [MeV] and becomes slow at $\sim 192.6$ [MeV]. We associate
a
critical temperature ``$T_c$'' $\sim 187\pm0.5$ [MeV]   to this
crossover phenomenon for comparison with extrapolated values of $T_c$ in chiral
perturbation theory (in our case $T_c$ is
localized as the point of inflection in the crossover
curve). Thus $T_c$ has increased by 5.3 \% compared to the chiral limit. This
result
is in agreement with the estimate of Leutwyler~\cite{leu}, who predicts $\Delta
T_c/T_c\sim 5\%$ if realistic pion masses are substituted for the chiral limit
with
$m_\pi=0$. When the critical temperature is extrapolated in the framework of
chiral
perturbation theory, the inclusion of finite quark masses delays the melting of
$\la\bar q q\r$ by $\Delta T\sim 20$ [MeV]~\cite{ger}, while the inclusion of
heavier
mesons in a dilute gas approximation has the opposite  effect. At finite quark
masses
it accelerates the melting from $T_c\sim 240$ [MeV] to 190 [MeV]. The
delay in the melting due to finite quark masses is intimately related to the
size of
the latent heat. The relation is revealed in the derivation of a
Clausius-Clapeyron-equation for QCD, cf.~\cite{leu}. Thus the
$T_c(m)$-dependence is
conclusive for the strength of a first order chiral transition.\\

{\it The case of realistic mass splitting, $\ve_0^{crit}/\ve_8^{crit}=0.77$.}
For
the realistic mass splitting induced by $m_s/m_{u,d}=18.2$ we find for the
critical
field strengths in large-$N_f$
\be\label{cv}
\ve_0^{crit}\leq(7\pm2)\cdot10^{-3} {\rm [GeV]}^3,\ \ \ \
\ve_8^{crit}\leq(-9.09\pm2.6)
\cdot10^{-3} {\rm [GeV]}^3
\ee
with $m_\pi=68.7,\  m_K=276.8, \ m_\eta=313.4,\
m_{\eta'}=928.6,\  m_{\sigma_\pi}=887.9, \ m_{\sigma_K}=890.9,\
m_{\sigma_\eta}=924.3, \ m_{\sigma_{\eta'}}=698.2$ (all masses in units of
[MeV]) or an
average pseudoscalar octet mass of $\la m_{ps}\r=203 {\rm [MeV]}$. The critical
values for the current quark masses are $m^{crit}_{u,d}=2.96\pm0.85{\rm [MeV]}$
and
$m_s^{crit}=54\pm15.4$ [MeV]. The reason why we give an upper bound on
$\ve^{crit}_{0,8}$ rather than precise values for the first order transition
boundary
is the same as in the $SU(3)$-symmetric case. The bound on  $\ve^{crit}_{0,8}$
could
still be improved by measuring data between $\ve_0=4\cdot10^{-3}$ [GeV]$^3$,
$\ve_8=-5.2\cdot10^{-3}$
[GeV]$^3$, where the  transition is still of first order, and the above values
for
$\ve^{crit}_{0,8}$. It should be remarked that only a rather fine resolution of
the $\la \bar qq\r / \la \bar qq \r_0 (T)$ curve in steps of $\Delta T \leq
0.1$
 [MeV] has revealed the true 1st order nature of the transition for $\ve_0 =
5\cdot 10^{-3}$ and $\ve_8 = -6.5\cdot 10^{-3}$ [GeV]$^3$ due to the small
gap in the condensate. The resolution in units of $\Delta T = 1$ [MeV] had
suggested
a smooth crossover behaviour already for these smaller values of $\ve_{0,8}$.

Thus we see that also in a large $N_f$-expansion tiny values for the quark
masses are
sufficient to weaken the chiral  transition and turn it finally into a
crossover
phenomenon. Similarly
tiny current quark masses are sufficient to eliminate the chiral transition in
an
$SU(3)$-Nambu-Jona-Lasinio model as a function of temperature and nuclear
density~\cite{lutz}.\\
 The
large-$N_f$-results for $\ve^{crit}_{0,8}$ are clearly
above  the
mean-field values of Eq. (\ref{ck}).
The result is plausible  as our saddle-point approximation goes beyond the
mean-field
calculation. The leading term in a $1/N_f$-expansion corresponds to the
summation of a class of diagrams, called ``super-daisies''~\cite{jai}.
Super-daisies have been summed up by Dolan and Jackiw~\cite{dol} to
circumvent the IR-divergence problem in an $N$-component $\phi^4$-theory.
Our application of the $1/N_F$-method to the linear sigma-model has been
similar in spirit. Certainly we cannot claim that the fluctuations we have
included so
far are representative for all fluctuations. In fact, the classical cubic term
of our
potential may still dominate the driving mechanism for the first order
transition below
the critical field strengths. Only the lattice calculation includes all
fluctuations
by simulating the full partition function at once (at least in principle). This
may
explain, why a discrepancy of $\approx 0.25$ remains
between the large-$N_f$-ratio and the lattice result for
$m^{crit}_{u,d}/m_{u,d}$.
We should keep in
mind, however, that the lattice result itself is not yet reliable, as we have
argued
in section 1.

The next question which arises in a comparison with the mean-field calculation
concerns the critical renormalized masses which should vanish for the critical
field
strengths at a second order transition. In the mean-field calculation
$m^2_{\sigma_{\eta'}}$ vanishes by construction for the set of critical
parameters,
and it is easily checked that $m_{\sigma_{\eta'}}$ is the only mass that
vanishes at
criticality. Our value for $m_{\sigma_{\eta'}}=749.5$ [MeV] is the tree level
input
mass  at zero temperature, which need not  vanish. The effective masses $X_Q$
are
temperature dependent. They slowly de/increase with $T$ as we will see in
section 6.
We have mentioned in the discussion of
the chiral limit (cf. section 4) that $X_Q, Q=1...4$ violate Goldstone's
theorem, if
they coincide with a dynamical mass in our approximation. Similarly it is here
not
obvious that one of these masses should induce an infinite correlation length
as $T$
approaches $T_c$. A careful renormalization prescription should be imposed to
identify the renormalized mass(es) that go to zero for critical  external
fields. It
would further allow an identification of the universality class of the
$SU(3)\times
SU(3)$-linear sigma-model for critical parameters
$\mu^2_0,f_1,f_2,g,\ve_0,\ve_8$.
We will investigate these questions in a forthcoming work.

{\it Critical meson masses for $m_s=0$.} Here the large-$N_f$-results are even
somewhat
smaller than  the mean-field results (cf. Eqs. (\ref{cs})). We find
\be\label{cw}
\ve_0^{crit}=(4\pm1)\cdot 10^{-4}{\rm [GeV]}^3\ \ \ \
\ve_8^{crit}=(3.06\pm0.76)\cdot10^{-4}
{\rm [GeV]}^3.
\ee
The associated average pseudoscalar octet mass is
 $\la m_{ps}\r=57.2\pm28.8{\rm [MeV]}$, and $m_{u,d}=1.7\pm0.4{\rm
[MeV]}$, while $m_s=0$ by construction.\\
The Columbia plot (cf. Fig.1) has suggested a concave shape of the first order
transition boundary. The three critical masses in mean-field are compatible
with
such a shape in an ($m_\pi , m_K$)- or an ($\hat m , m_s$)-diagram, although
one
should keep in mind the large error bars due to the ambiguity in the tree-level
parametrization of the sigma-model and the sensitive dependence of the boundary
on
the $\sigma_{\eta^{(')}}$-mass input. The three critical masses in large-$N_f$
do not confirm the conjectural concave shape. Only for a realistic ratio of
$\ve_0/\ve_8$
the large-$N_f$-result lies clearly above the mean-field value for the critical
masses.
The error here is at least as large as in the mean-field case. The error could
have been
further reduced, but the remaining size would reflect the difficulty in
disentangling
a very weak first order transition from a rapid crossover phenomenon as we have
mentioned
above.

In the next section we will see the change in the  crossover behaviour, as
$\ve_0,\ve_8$ are further increased to induce realistic mass values.


\section{The realistic mass point}

If we use for $\mu^2_0,f_1,f_2,g$ the values of the chiral limit and choose
$\ve_0=0.0265$~[GeV]$^3$, $\ve_8=-0.0345$~[GeV]$^3$, Eqs.~(\ref{aj}-\ref{ak})
lead to
(pseudo)scalar meson masses which are listed in Table~1. A comparison to the
experimental values shows reasonable agreement for the pseudoscalar mesons.
Therefore
we call this  point the ``realistic'' mass point. The experimental values which
are associated to the scalar meson masses depend on the identification, which
is
indicated in a separate row of Table~1. The mass splitting between $K^*_0$ and
$a_0$
comes out too small in our case. We could have further optimized our choice of
$\ve_0$
and $\ve_8$ to improve the agreement with the experimental mass values, but
such an
optimization should be inconsequential for our results.
\begin{table}[htb]
\begin{center}
\caption{Tree level parametrization of the
$SU(3)\times SU(3)$ linear
sigma-model for the realistic mass point}
\begin{tabular}{|l|l|l|l|l|l|l|}
\multicolumn{7}{c}{Input} \\
\multicolumn{7}{c}{ }\\
\hline
$\mu^2_0$ [GeV]$^2$& $f_1$ & $f_2$ &
$g$ [MeV]&
$f_{\pi}$ [MeV] & $\ve_0$ [GeV]$^3$ & $\ve_8$ [GeV]$^3$\\
\hline
$5.96\cdot10^{-2}$ & 4.17 & 4.48 & -1,812.0 & 94 & 0.0265 & -0.0345\\
\hline
\end{tabular}
\begin{tabular}{|l|l|l|l|l|l|l|l|l|}
\multicolumn{9}{c}{} \\
\multicolumn{9}{c}{Output (all masses are understood in units of [MeV])} \\
\multicolumn{9}{c}{ }\\
\hline
& $m_\pi$ & $m_K$ & $m_\eta$ & $m_{\eta'}$ &
$m_{\sigma_\pi}$ & $m_{\sigma_K}$ & $m_{\sigma_\eta}$ & $m_{\sigma_{\eta'}}$\\
\hline
real. & & & & & & & &\\
mass  & 129.3 & 490.7 & 544.7 & 1045.5 & 1011.6 & 1031.2 & 1198.0 & 749.5\\
point & & & & & & & & \\
\hline
exp. & & & & & & & &\\
mass  & 138.0 & 495.7 & 547.5 & 957.8 & 980 if & 1322.0 if & 1476.0 if & 975
if\\
values  & & & & & $\sigma_\pi\equiv$ & $ \sigma_K\equiv$ &
$\sigma_\eta\equiv$ & $\sigma_{\eta'}\equiv$\\
        & & & & & $a_0$ & $K_0^*$ & $f_0 (1476)$ & $f_0 (975)$ \\
\hline
\end{tabular}
\end{center}
\end{table}

{\it Crossover in the condensates.} The crossover behaviour for the normalized
light and strange
quark condensates are displayed in Fig.~8.
\begin{figure}[tbh]
\vspace{13.0cm}
\caption{Light ($\la\bar qq\r$) and strange ($\la\bar ss\r$) quark condensates
normalized to their corresponding values at zero temperature as a function of
temperature. The crossover behaviour is   most rapid between $181.5\leq
T\leq192.6$~[MeV].}
\end{figure}
The rapid crossover leads to a decrease of $\sim50$~\% of the condensate at
zero
temperature $\la\bar qq\r_0$ in $\la\bar qq\r_T$ over a temperature interval
$\Delta
T\sim10$~[MeV], while $\la\bar ss\r_T$ stays remarkably constant up to
 $T\sim 197$~[MeV]
where it starts to decrease rather slowly. The physical reason is obvious. It
is harder to
excite mesons with strange quarks than with light quarks. The same qualitative
behaviour of $\la\bar ss\r_T$ has been noticed by Hatsuda and Kunihiro
\cite{hat} in a $U_{N_f}(3)$-version of the Nambu-Jona-Lasinio model.
Also the location of the crossover region in the NJL-model is around
$T\sim200$~[MeV].\\
We have indicated  the error bars only in  the crossover region where
they are largest. As
outlined in section~3, their main sources are the uncertainty in precisely
locating
the saddle point value $sad^*$ and the minima
$\la\sigma_0\r,\la\sigma_8\r$ of $\hat U_{eff}$. When the errors in the
current quark masses are added, which enter Eq.~(14), we obtain  $\Delta\la\bar
qq\r\sim3.1\cdot10^{-3}$~[GeV]$^3,\Delta\la\bar
ss\r\sim6.7\cdot10^{-3}$~[GeV]$^3$
 at $T = 180$~[MeV] in the transition region.
In Fig.~8 we have indicated only the numerical errors, the contribution from
the
current quark masses has been left out. Compared to critical meson masses
$(\ve_0^{crit}=7\cdot10^{-3}$~[GeV]$^3,\ve_8^{crit}=-9.1\cdot10^{-3}$~[GeV]$^3$),
 the crossover happens over a
larger temperature interval. We find $\Delta [\la\bar qq\r_T/\la\bar
qq\r_0 ]^{real} / \Delta [\la\bar qq\r_T/\la\bar
qq\r_0 ]^{crit}
\sim 52$~\% if $\Delta [\la\bar qq\r_T / \la\bar qq\r_0 ]$ denotes the
normalized
condensate change per
1~[MeV] temperature interval in the rapid part of the crossover region.
Nevertheless the crossover in the quark condensate $\la\bar qq\r$ seems
to be sharp even for realistic masses. Such a rapid change may lead to visible
changes
in hadron masses which depend on temperature and condensates. It could be
manifest in
hadronic or dilepton spectra in heavy-ion experiments.
\begin{figure}[htb]
\vspace{13.0cm}
\caption{Contour plot of $\hat U_{eff}(\sigma_0,\sigma_8)$ in a projection on
the
$(\sigma_0,\sigma_8)$-plane for four temperatures. In Fig. 9a)  $(T=10$ [MeV])
we see
two separated minima (indicated with crosses), one unphysical with
$\sigma_0<0,\sigma_8<0$ and the absolute physcial minimum with
$\sigma_0>0,\sigma_8<0$. In the crossover region at $T=190$ [MeV] (Fig. 9b))
both
minima lie in the physical sector with $\sigma_0>0,\sigma_8<0$. The minima
coincide
for temperatures $\geq 200$ [MeV] (Fig. 9c) with $T=200$ [MeV]). For increasing
temperature the one remaining minimum moves inwards, but stays away from zero
due to
$\ve_{0,8}\not=0$, as is seen in Fig. 9d) for $T=350$ [MeV].}
\end{figure}

The crossover behaviour in the condensates is also revealed in two-dimensional
contour plots of the effective potential $\hat
U_{eff}(\sigma_0,\sigma_8;sad^*)$. In
Fig. 9 we show equipotential lines for four temperatures in a projection of
$\hat
U_{eff}$ on the $(\sigma_0,\sigma_8)$-plane. In Fig. 9a) we see two separated
local
minima for $T=10$ [MeV], one in the sector $\sigma_0<0,\sigma_8<0$ and another
one
in the physical sector $\sigma_0\geq0,\sigma_8\leq 0$. (In the chiral limit we
would
find three minima due to $\ve_0=0=\ve_8$.) In the crossover region at $T=190$
[MeV]
(Fig. 9b)) we still have two separated minima, where the left one is a local
minimum
and the right one is the absolute minimum of $\hat U_{eff}$. Now both minima
lie in
the physical sector. At $T=200$ [MeV] (Fig. 9c))  the minima have merged in  a
single one. As the temperature is further increased, the absolute minimum moves
inwards,  but stays away from zero due to the explicit symmetry breaking fields
$\ve_0$ and $\ve_8$ (cf. Fig. 9d) for $T=350$ [MeV]). The elliptical shape of
the
contour plot reflects the violation of spherical symmetry in
$(\sigma_0,\sigma_8)$-space due to the explicit symmetry breaking, the cubic
term
in the classical potential and the thermal part. As the  thermal part gets more
dominant at higher temperatures, its high-temperature limit changes the
elliptical
shape to an approximately spherical symmetric shape. The two minima below 200
[MeV]
may be easily misinterpreted as the coexistence of the symmetric and the chiral
phase~\cite{met}, if one does not realize that the minimum
$(\sigma_0,\sigma_8)$
on the r.h.s. of Fig. 9b) stays the absolute minimum for all temperatures.

{\it Thermodynamics.} Further characteristics of the crossover phenomenon are
the
variations in the energy and entropy densities over the temperature interval in
the
crossover region. At this place one should recall the very definition of a
first
order phase transition. At the transition point {\it at least one} of the first
derivatives of a suitable thermodynamic potential should behave discontinuously
in
the infinite volume limit. Thus a crossover in the condensates in general
does not exclude a
finite gap in the energy or entropy densities.
\begin{figure}[thb]
\vspace{13.0cm}
\caption{Entropy density $s$ over $T^3$, energy density $\ve$ over $T^4$ and
pressure
$p$ over $T^4$ in the large $N_f$-expansion for the realistic mass point.
Errors are
only indicated for $s/T^3$.}
\end{figure}
In Fig.~10 we have plotted $s/T^3,\ve/T^4$, and $p/T^4$ as calculated according
to
Eqs.~(\ref{bn}-\ref{bp}).
The data points  are strongly
fluctuating within a range which is indicated by the error bars for the entropy
curve. The errors for the energy density  are of similar size.
Errors enter via the effective masses $X^2_Q$,
which depend on $\la\sigma_0\r,\la\sigma_8\r$ and $sad^*$. We have used for
$\Delta\la\sigma_0\r\sim(1.0-3.0)\cdot10^{-3}$~[GeV]$ =
\Delta\la\sigma_8\r$ and
$\Delta\la sad^*\r\sim0.11$~[GeV]$^2$ in the crossover region between $T = 182-
193$~[MeV]. The pressure behaves
continuously as a function of $T$ if it is calculated as $p=(-\hat U_{eff})$.
There
is only a change of slope in the critical crossover region. A direct
calculation of $p$
with an integral expression pretends a discontinuity.  The $p/T^4$-curve in
Fig.~10
is obtained from $(p=-\hat U_{eff})$. As change in the entropy density we find
from the
actually measured data
\be \label{da}
T\cdot\Delta s\leq0.16\ {\rm [GeV/fm}^3] \ee
in a temperature interval $181.5\leq T\leq192.6$~[MeV], where $T\cdot\Delta s$
is
calculated as $T_2\cdot s(T_2)-T_1\cdot s(T_1)$. As rapid change in the energy
density
we find
\be \label{db}
\Delta\ve\leq0.13\ {\rm [GeV/fm}^3] \ee
or
$\Delta\ve/T^4_c\equiv\ve(T_2=192.6$~[MeV])$/T^4_2-\ve(T_1=181.5$~[MeV])$/T_1^4=0.29$
over the same temperature range. For comparison we mention that the gap in the
gluonic
energy density $\Delta\ve_g$ in a pure $SU(3)$ gauge theory leads
to~\cite{eng,bro}
\be \label{dbb}
\frac{\Delta\ve_g}{T^4_c}=\left\{ \begin{array}{ll}
2.44\pm0.24,&{\rm for}\ N_\tau=4\\ 1.80\pm0.18,&{\rm for}\ N_\tau=6,\\
\end{array}\right. \ee
where $N_\tau$ refers to the lattice extension in time direction. These values
are by
an order of magnitude larger than our value for the mesonic contribution
$\Delta\ve/T^4_c\sim0.25$, defined as indicated above. The decrease of
$\Delta\ve_g/T^4_c$ under an increase of the time extension $N_\tau$ indicates
strong
finite size effects. Going to larger lattices this tendency may continue and
further
reduce the latent heat, but it also gives us a warning. The contribution of
$\Delta\ve_g$ to the total energy gap may be superimposed on the slow change of
$\ve$
that we have found in the crossover region and make the crossover in the total
energy
density more rapid. If the size of the errors in
$sad^*, \la \sigma_0 \r,\la \sigma_8 \r$ is assumed as
above,
$s$ could vary in the crossover region between $T_1$ and
$T_2$ like
\be \label{dc}
s(T_2)\cdot T_2-s(T_1)\cdot T_1\leq0.18\ {\rm [GeV/fm}^3], \ee
where the resulting error $\Delta(s/T^3)$ has been estimated as $\pm0.46$, cf.
Fig.~10.
In an infinitesimally small temperature
interval such a gap in $s$ would lead to a finite latent heat of
$\leq0.2$~[GeV/fm$^3]$. Thus Eq.~(\ref{dc}) gives a loose upper bound on the
latent
heat which could be compatible with our data due to the large errors in the
crossover
region. The bound comes out even smaller if the error of $0.46$ is interpreted
as
$\Delta s/T^3_c$ with ``$T_c$''~=~187~[MeV]. It leads to $\Delta
L=T_c\cdot\Delta
s\leq0.074$~[GeV/fm$^3$]. Both bounds are even smaller than Leutwyler's value
of
0.4~[GeV/fm$^3]$ for $T\cdot\Delta s$ \cite{leu}, which has been obtained from
a
Clausius-Clapeyron equation in the framework of chiral perturbation theory. The
small
size of the latent heat is finally a consequence of the sensitivity of $T_c$ to
the
inclusion of finite quark masses.\\
In view of heavy-ion experiments there need not be a latent heat in the strict
sense, which occurs over a time period during the phase conversion, where the
temperature stays exactly constant. Multiplicity fluctuations in rapidity
distributions of
charged particles are also induced if $\Delta s$ is sufficiently large over a
small, but finite temperature interval (say of the order of $\sim10$~[MeV]).
In van Hove's formulation \cite{hov} the physical conditions of a
first order transition are identical to those of a rapid crossover phenomenon.

\indent Clausius-Clapeyron equations relate the
discontinuities in the condensate and the entropy/energy densities. Although
they
strictly apply to first order transitions in the form of
Eqs.~(\ref{dcc}-\ref{dccc})
below~\cite{leu}, we have tested the relations for our rapid crossover
phenomenon in two
forms. The first one is  \be \label{dcc} disc\la\bar
qq\r_T\Bigm\vert_{r.c.}=\frac{\Delta T_c}{\Delta\hat m}\Bigm\vert_{r.c.}disc\
s\Bigm\vert_{r.c.}, \ee where $disc...|_{r.c.}$ refers to the rapid change
(``discontinuity'') in the crossover region and $\Delta T_c$ is the change in
``$T_c$'' under a variation $\Delta\hat m$ of the current light quark masses.
The second
version is
 \be \label{dccc} \frac{\Delta
T_c}{T_c}\Bigm\vert_{r.c.}=\left|\frac{disc\la\bar qq\r_T}{\la\bar
qq\r_0}\right|_{r.c.}\left|\frac{f^2_\pi m^2_\pi}{\Delta\ve}\right|_{r.c.}. \ee
\noindent In
Eq.~(\ref{dcc}) we use $\Delta T_c/\Delta\hat m=(0.187-0.178)/(0.011),\Delta s$
=
0.0061 [GeV]$^3$, \\$disc\la\bar qq\r_T$ $=$ $0.5\cdot(0.22)^3$, and obtain
0.005
for the l.h.s. and for the r.h.s. of Eq.~(\ref{dcc}). In Eq.~(\ref{dccc}) we
have $\Delta T_c$ as above, $T_c=178$~[MeV], $disc\la\bar qq\r_T/\la\bar
qq\r_0=\la\bar qq\r_{T_1}/\la\bar qq\r_0-\la\bar qq\r_{T_2}/\la\bar qq\r_0=0.5,
f_\pi=94$~[MeV], $m_\pi=129$~[MeV] (the value of our realistic mass point), and
$\Delta\ve=0.001$~[GeV]$^4$. This way we obtain for the r.h.s. 7 \% and for the
l.h.s. 5 \% in Eq.~(\ref{dccc}). The agreement of the order of magnitude on
both
sides indicates that the crossover phenomenon is still rapid enough to satisfy
analogous relations to a first order transition in the strict sense. \\
The same relations have been checked for the first order transition in the
chiral limit.
 Eq.s (64),(65) with the appropriate chiral input data predict for $\Delta \ve
/T^4_c
\sim 1.0 \pm 0.2 \sim \Delta s /T^3_c$. The actually measured gap $\Delta
s/T^3_c$
in chiral thermodynamics comes out as $0.5\pm0.2$. An error of $0.2$ is easily
induced
by an error of $10$~[MeV] in the condensates $\la \sigma_0 \r,\la \sigma_8 \r$
at
the phase transition. Thus the relations are approximately satisfied also in
the
chiral limit
within the relatively large errors.

We
have further analyzed the contribution of the strange (pseudo)scalar mesons $K$
and
$\sigma_K$ to the total  energy density $\ve$. Their contribution can be
completely
neglected below 40~[MeV]. It increases with temperature to $\sim31$~\%  in the
crossover region around $T=187$~[MeV]. After the crossover the tendency
continues, but
is no longer conclusive for us due to the lack of quark degrees of freedom in
the
chiral symmetric phase.\\
The strangeness content of the plasma has been estimated in a
lattice simulation~\cite{kog} with light quark masses  of $m_u/T=m_d/T=0.05$
and one heavier quark mass of $m_s/T=1.0$. In the transition region one finds
for the
ratio of the fermionic energy densities $\ve_F(m_s/T=1)/\ve_F(m_u/T=0.05)\simeq
0.5$.
The good agreement with our ratio of mesonic contributions from strange and
non-strange
(pseudo)scalar mesons ($\ve$(strange mesons)/$\ve$(non-strange mesons))$\sim
0.45$ may
be accidental, because the lattice estimate is based on perturbative relations
for
the energy density. A fully non-perturbative lattice calculation along the
lines of
Engels et al.~\cite{eng} in a pure $SU(3)$ gauge theory is still outstanding if
fermions are included. Nevertheless, it should be mentioned that an
extrapolation for
fermionic energy densities $\ve_F(m/T=1)$ and $\ve_F(m/T=0)$ from the plasma
phase to
the transition region under the assumption of ideal fermion gases leads to
$\ve_F(m/T=1)/\ve_F(m/T=0)=0.88$~\cite{kars}, which is twice the amount of our
ratio for mesons in the transition region. This may be taken as an indication
that
the strangeness contribution to $\ve$ around $T_c$ cannot be derived from an
underlying ideal gas of quarks of $m_s$ and $m_{u,d}$-quarks.

{\it Temperature dependence of effective masses.} Fig.~11 displays the
temperature
dependence of the effective masses $X_Q,\ Q=1,...,8$ up to temperatures of
300~[MeV].
\begin{figure}[htb]
\vspace{13.0cm}
\caption{Temperature dependence of the effective mass squares $X_Q,$ $Q=1$,
$...,8$
for realistic meson masses. From bottom to top:
$X_\pi,X_K,X_\eta$, $X_{\sigma_{\eta'}},X_{\sigma_\pi},X_{\sigma_K}$,
$X_{\eta'},
X_{\sigma_\eta}$.}
\end{figure}
We have found a degeneracy of all (pseudo)scalar meson masses at higher
temperatures
($T\geq400$~[MeV]), where the sigma-model fails to describe the plasma phase.
Immediately above the rapid crossover region chiral symmetry is partially
restored. The restoration in strange and non-strange sectors is achieved quite
differently. We find an approximate degeneracy between $X_\pi$ and
$X_{\sigma_{\eta'}},\ X_K, X_\eta$ and $X_{\sigma_K}, X_{\eta'}$ and
$X_{\sigma_\pi}$,
while $X_{\sigma_\eta}$ becomes degenerate with other masses only at
$T\sim400$~[MeV].
The $\sigma_\pi$-mass never meets the $\pi$-mass, because the $U_A(1)$-anomaly
term
(proportional to $g$) enters the mass formulas with opposite sign. From lattice
results it is very likely that other modes than $X_\pi,X_{\sigma_{\eta'}}$
dissolve
in their constituents and cease to exist in the chiral symmetric phase. Coming
from
the broken phase we notice that the effective mass $X_{\sigma_{\eta'}}$ drops
to a
remarkably small value in the crossover region $(X_{\sigma_{\eta'}}\sim(X_\pi$
at
$T=0)\sim130$~[MeV]), while $X_\pi$ is almost zero. Furthermore, we have
displayed the errors in $X_\pi$ and $X_{\sigma_{\eta'}}$ in the transition
region.
The errors have increased with $T$ to these values
 due to the contribution of $\Delta sad(T)$. For $X_\pi$ the error is
particularly large in the crossover region, since the error $\Delta X_\pi$ is
proportional to $1/X_\pi$ and $X_\pi$ is almost zero in this region. Before we
jump to
conclusions about measurable effects due to tiny $\sigma_{\eta'}$ and
$\pi$-masses in
the crossover region, we should clarify the precise physical meaning of the
effective
masses $X_Q$. We recall that the main contribution to the temperature
dependence of
$X_Q$ comes from the saddle point contribution $sad^*(T)$, which is just the
leading
term in the $1/N_f$-expansion.


\section{Summary of results and conclusions}

In agreement with the general expectation results of
the high-$T$-expansion   are qualitatively correct, but  fail quantitatively.
The
critical temperature in the chiral limit deviates by $\sim 85$ [MeV] from $T_c$
in the
numerical  evaluation, which is applicable also in the  transition/crossover
region.
The crossover region for realistic meson masses is shifted by roughly $80$
[MeV]
between the high-$T$- and the numerical results.
In our model the high-$T$-expansion practically never reproduces the numerical
results in a quantitative way. While the meson condensate $\la \sigma_0 \r$ in
the realistic mass case has dropped to values $\leq 2$~[MeV] around $T\sim
400$~[MeV]
in the high-$T$-expansion, it is still larger than $30$~[MeV] at this
temperature
in the numerical calculation. The temperature lies already far outside the
applicability
range of the model.

\indent The crossover in the light quark condensate looks still rapid for
realistic
meson masses $(\Delta(\la\bar q q\r_T)\sim50\%$ of $\la\bar q q\r_0$ in a
temperature
interval of 10 [MeV] in the crossover region). Less rapid look the variations
in the
energy and entropy densities between 181.5 [MeV] $\leq T\leq 192.6$ [MeV]. We
find
$\Delta\ve\sim 0.13\pm0.02$ [GeV/fm$^3]$ and $T\cdot\Delta_s\sim 0.16\pm0.02$
[GeV/fm$^3]$.
As a loose bound on the latent heat we obtain $0.2$~[GeV/fm$^3]$, as a more
stringent one $0.1$ [GeV/fm$^3]$, if the
change in $\ve$ and $s$ would occur over an arbitrarily small temperature
interval.
The main contributions to the numerical errors which
prevent us from a unique interpretation of the transition/crossover
region come from the uncertainties in the
saddle point value and the  minima $\la\sigma_0\r,\ \la\sigma_8\r$ of $\hat
U_{eff}$.

For temperatures above 116~[MeV] the saddle point variable gives a positive
contribution to the effective masses entering the argument of the logarithm in
$\hat
U_{eff}$. It increases with temperature. This was one of the reasons why we
have
chosen the large-$N_f$-expansion. The original hope to completely avoid
imaginary
parts of $\hat U_{eff}$ in this scheme could not be confirmed by the results.
The
effective potential is still plagued with imaginary parts for certain regions,
where
$|\sigma_{0,8}|\leq|\la\sigma_{0,8}\r|$ and over the entire temperature range
we
have studied  (up to 250~[MeV] and above).\\
The large error bars on $\ve$ and $s$ in the transition region may leave some
doubts
on the smooth nature of the conversion between the chiral symmetric and the
chirally
broken phase, in particular, because  a smooth crossover in the condensates
does not
automatically imply a smooth change in $\ve$ and $s$. Even if the transition
would
be of  first order, and
even if we use a loose bound on the latent heat, it is as small as 0.2
[GeV/fm$^3]$.

The ratio of critical to realistic current light quark masses has been
estimated as
$m^{crit}_{u,d}/m_{u,d} \sim 0.03 \pm 0.02$ in mean-field and as
$m^{crit}_{u,d}/m_{u,d}
\sim 0.26\pm0.08$ in large-$N_f$. The large-$N_f$ result implies that realistic
quark/meson masses lie less deeply in the crossover region than the mean-field
result
suggests. Due to the fluctuations which are effectively included in the
large-$N_f$-
approximation the ratio in large-$N_f$ is about half of the lattice result.
The $average$ pseudoscalar octet masses are 126.4 [MeV] and 203 [MeV] for the
realistic ratio of $\ve_0^{crit}/\ve_8^{crit}$ in mean-field and large-$N_f$,
respectively. Note that an error of 30 \% in the critical quark masses blows up
to an error of 55 \% in the critical meson masses, if the meson mass squares
are linear in the current quark masses. Thus the critical values for the meson
masses
 are not very conclusive.

In view of heavy-ion experiments one may conclude that there is still little
hope for
measurable experimental signatures of the crossover region, although the
ratio $m^{crit}_{u,d}/m_{u,d}$ has become larger than the mean-field estimate.
 Quantities which are
indirectly observable such as condensates, energy and entropy densities vary
too
smoothly during the phase conversion. Experimentalists should be  warned,
however, to
accept these conclusions without care.

The tendency that $m^{crit}_{u,d}/m_{u,d}$ is larger in large-$N_f$ than in
mean-field
is plausible as an effect of the included fluctuations, but  as we have seen
 this effect could
not be confirmed for the $SU(3)$-symmetric or the $m_s=0$ case. The critical
mass ratios there were even smaller in large-$N_f$ than in mean-field or at
least
of the same order of magnitude within the errors.
This gives a hint that the fluctuations we include in our approximation are
likely not the only important ones. In particular, it is not clear that they
account for fluctuations which induce a renormalization of the quartic and
cubic couplings in the Lagrangian.

Recently it has been raised by Gavin, Goksch and Pisarski~\cite{gav} that the
first
order of the chiral transition may be mainly $fluctuation\  induced$. A
fluctuation
induced transition has been first discovered by Coleman and Weinberg
\cite{cole}. It
refers to the situation that a system with more than one relevant coupling can
have
a first order transition induced by quantum fluctuations. In an understanding
of the
renormalization group the fluctuation induced transition occurs due to
so-called
run-away RG trajectories (see e.g. \cite{shen}). Such an origin for a first
order
transition has been demonstrated in $4-\ve$ dimensions for the $SU(N)\times
SU(N)$
linear sigma model by Paterson \cite{pat}, for more general symmetry groups
$(O(N)\times O(M),U(N)\times U(M))$ by Pisarski and Stein \cite{pisa}. In 4
dimensions Shen \cite{shen} has shown in a numerical simulation of a
$U(N)\times
U(N)$ symmetric scalar model that a fluctuation induced first order transition
occurs in particular for large coupling $f_2$. For large and slowly varying
$f_2,\
f_1$ runs fast from large to small values as a function of energy scale.
The same tendency in an $SU(3)\times SU(3)$ linear sigma model would be
compatible with a larger critical mass ratio $m^{crit}_{u,d}/m_{u,d}$. The
larger mass ratio of the lattice simulation points into the same direction,
since
 the lattice includes all fluctuations at once.\\
In our present approach it is difficult to disentangle the driving mechanism
for the
first order transition below the critical mass values. A further improvement by
including
subleading corrections in $1/N_f$ or an $\ve$-expansion in $d=4-\ve$ dimensions
should clarify, whether the ratio $m^{crit}_{u,d}/m_{u,d}$ does change to
larger values. If the realistic finite quark masses lie in fact closer to the
first
order transition boundary than our results suggest, remnants of the first or
second
order transition may be more easily visible in experiments.

A second warning should be mentioned not to take the smooth crossover for
guaranteed
by the present approach. So far we have completely neglected the quark- and
gluonic
substructure. In particular the rearrangement of gluonic degrees  of freedom
has not
been taken into account. The gluonic contribution to the change in entropy and
energy densities may well accelerate  the crossover process. Note that we have
chosen
the couplings in the sigma-model as temperature and energy scale independent
over all
temperatures up to $T_c$. In principle, the temperature and scale dependence of
the couplings should be derived from QCD rather than being assumed. Less
ambitious it
may be derived from an effective model underlying the sigma-model and
containing quark
and gluonic degrees of freedom. For the critical mass ratio this offers at
least a
chance for being closer to 1. Work in this direction is in progress.

\section*{Acknowledgments}
We would like to thank H.-J. Pirner for useful discussions.

\end{document}